\documentclass[%
superscriptaddress,
groupedaddress,
unsortedaddress,
altaffillsymbol,
reprint,
 amsmath,amssymb,
 aps,
]{revtex4-2}

\usepackage{graphicx}
\usepackage{dcolumn}
\usepackage{bm}

\begin{document}

\preprint{APS/123-QED}

\title{$\beta^-$ decay of neutron-rich $^{45}$Cl at magic number N=28}

\author{Soumik Bhattacharya$^{1}$\footnote{Corresponding author: soumik.kgpiit@gmail.com}} 
\author{Vandana Tripathi$^{1}$\footnote{Corresponding author: vtripath@fsu.edu}} 
\author{S.~L.~Tabor$^{1}$}
\author{A.~Volya$^{1}$}
\author{P.~C.~Bender$^2$}
\author{C.~Benetti$^{1}$} 
\author{M.~P.~Carpenter$^{3}$} 
\author{J.~J.~Carroll$^4$}
\author{A.~Chester$^{5,6}$}
\author{C.~J.~Chiara$^4$}
\author{K.~Childers$^{5,7}$}
\author{B.~R.~Clark$^{8}$}
\author{B.~P.~Crider$^8$}
\author{J.~T.~Harke$^9$}
\author{S.~N.~Liddick$^{5,6,7}$}
\author{R.~S.~Lubna$^6$}
\author{S.~Luitel$^8$}
\author{B.~Longfellow$^{5,10}$}
\author{M.~J.~Mogannam$^{5}$}
\author{T.~H.~Ogunbeku$^{5,6,8}$}
\author{J.~Perello$^{1}$} 
\author{A.~L.~Richard$^{5,9}$}
\author{E.~Rubino$^{1}$} 
\author{S.~Saha$^2$}
\author{O.~A.~Shehu$^8$}
\author{R.~Unz$^8$}
\author{Y.~Xiao$^{5,8}$}
\author{Yiyi Zhu$^2$}

\affiliation{$^{1}$Department of Physics, Florida State University, Tallahassee, Florida 32306, USA}
\affiliation{$^2$Department of Physics, University of Massachusetts Lowell, Lowell, Massachusetts 01854, USA}
\affiliation{$^{3}$Physics division, Argonne National Laboratory, Argonne, Illinois 60439, USA}
\affiliation{$^4$U.S. Army Combat Capabilities Development Command Army Research Laboratory, Adelphi, Maryland 20783, USA}
\affiliation{$^5$National Superconducting Cyclotron Laboratory, Michigan State University, East Lansing, Michigan 48824, USA}
\affiliation{$^{6}$Facility for Rare Isotope Beams, Michigan State University, East Lansing, Michigan 48824, USA}
\affiliation{$^{7}$Department of Chemistry, Michigan State University, East Lansing, Michigan 48824, USA}
\affiliation{$^{8}$Department of Physics and Astronomy, Mississippi State University, Mississippi State, Mississippi 39762, USA}
\affiliation{$^{9}$Lawrence Livermore National Laboratory, Livermore, California 94550, USA}
\affiliation{$^{10}$Department of Physics and Astronomy, Michigan State University, East Lansing, Michigan 48824, USA}

\date{\today}

\begin{abstract}

Results from the study of $\beta^-$-decay of $^{45}$Cl, produced in the fragmentation
of a 140-MeV/u $^{48}$Ca beam, are presented. The half-life for $^{45}$Cl $\beta$-decay 
is measured to be 513(36) ms. The $\beta^-$ and $\beta^- 1n$ decay of $^{45}$Cl populated 
excited states in $^{45,44}$Ar, respectively. On the basis of $\gamma$-ray singles and 
$\gamma$-$\gamma$ coincidence data, decay schemes for the two daughter nuclei have been 
established. They are compared with shell model calculations using the FSU 
interaction. The low-lying negative parity states for $^{45}$Ar are well described by a
single particle (neutron) occupying orbitals near the Fermi surface, whereas neutron 
excitations across the $N = 20$ shell gap are needed to explain the positive-parity
states which are expected to be populated in allowed Gamow-Teller $\beta$-decay of $^{45}$Cl.
The highest $\beta$-feeding to the 5/2$^+$ state in $^{45}$Ar from the ground state of $^{45}$Cl
points towards a 3/2$^+$ spin-parity assignment of the ground state of the parent over
the other possibility of 1/2$^+$. 
The high Q$_{\beta^-}$ value of $^{45}$Cl decay allows for the 
population of $1p1h$ states above the neutron separation energy in $^{45}$Ar leading to
positive parity states of $^{44}$Ar being populated by removal of one neutron from the 
 $sd$ shell. The spin-parities of the excited levels in $^{44}$Ar are tentatively assigned 
 for the first time by comparison with the shell model calculations. The 2978~keV level of 
 $^{44}$Ar is identified as the excited 0$^+$ level which could correspond to a different 
 configuration from the ground state.

\end{abstract}

\maketitle


\newpage

\section{\label{sec:intro}Introduction}

In the past few decades a primary focus of nuclear structure studies has been understanding whether 
the known magic numbers that appear to hold good near stability, remain so as the drip lines 
are approached \cite{Warner_magic_number,BROWN2001,Sorlin_newmagicnumbers} where the 
proton-neutron asymmetry is large. The magic numbers $N=28$ and $Z=28$ are the lowest ones whose 
emergence requires a strong spin-orbit interaction and thus are of particular interest for the 
experimental and theoretical study of exotic nuclei far from stability to understand the 
isospin dependence of the spin-orbit interaction. There are several examples of experimental 
evidence, accompanied by theoretical calculations, which indicate that the $N=28$ shell gap below 
$^{48}$Ca reduces continuously with decreasing proton number. 
Just two protons away from $^{48}$Ca, 
the excitation energy of first 2$^+$ state in $^{46}$Ar drops considerably~\cite{46Ar_Calinescu}.
Further, around doubly magic $^{48}$Ca which is considered spherical the nuclear 
shape changes rather rapidly to develop deformation in $^{42}$Si~\cite{42Si_Bastin}
while shape coexistence is observed in 
$^{44}S$~\cite{44S_Force2010,44S_isomer_Parker,shell_evolution,Glasmacher_44S,44S_shape_warner,longfellow_44S_21}.

\begin{figure}
	\includegraphics[width=\columnwidth]{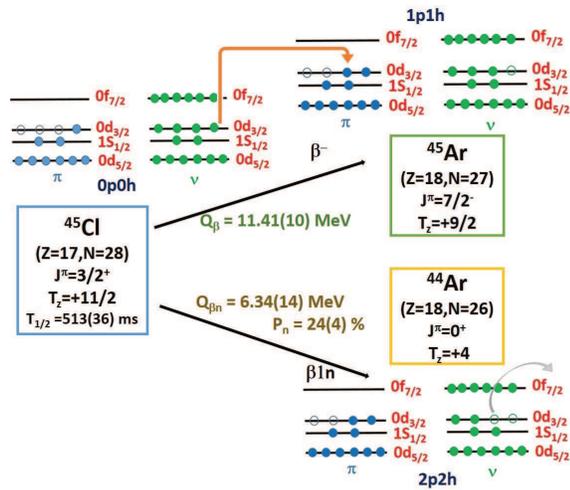}
	\caption{\label{fig:45Cl_decay_pattern}
	Schematic representation of $^{45}$Cl decay via $\beta 0n$ and $\beta 1n$ channels. 
 The boxes display the 
 ground-state spin-parity assignments of the corresponding nuclei. The dominant neutron and proton 
 configuration for the ground state of parent nucleus $^{45}$Cl and the 
 excited states expected to be populated in $^{45}$Ar and $^{44}$Ar by $\beta$ and $\beta 1n$ 
 are shown. The arrows show the transformation of neutron from $^{45}$Cl to proton and removal of 
 neutron from $^{45}$Ar involving the possible orbitals to produce excited states of $^{45}$Ar 
 and $^{44}$Ar, respectively.
	}
\end{figure}

The study of the excitation of protons in odd-$Z$ nuclei through measurements of excited states 
in the K, Cl and P isotopes have indicated a near-degeneracy of the proton $d_{3/2}$ and 
$s_{1/2}$ orbitals approaching $N = 28$ \cite{Gade_K_Cl_P_ProtonOrbital}. The increase of 
collectivity away from $Z=20$ as well as the degeneracy of $\pi$$d_{3/2}$ and $\pi$$s_{1/2}$ 
orbitals can be explained by the
monopole part of the tensor force~\cite{Otsuka_tensor_2005}, which is attractive
between the $\nu$$f_{7/2}$ and $\pi$$d_{3/2}$. With the degeneracy of the two proton orbitals, 
the ground states of odd-$A$ Cl isotopes are found to vary between 3/2$^+$ and 1/2$^+$.
The ground state spin-parity of $^{45}$Cl is not known experimentally, though both
1/2$^+$~\cite{Gade_K_Cl_P_ProtonOrbital} and 3/2$^+$~\cite{stroberg} are predicted
as possible spins based on different calculations. The ground states of odd-$A$ 
$^{37-45}$Ar, ($Z=18$) isotopes on the other hand are anticipated to be 7/2$^-$ in a simple 
filling of the orbitals due to a neutron hole in the $f_{7/2}$ orbital. 
However, the ground state spin/parity changes between 5/2$^-$ and 7/2$^-$ throughout 
the Ar isotopic chain from $N=20$ to $N=28$. A charge radius measurement using laser spectroscopy 
has found the ground state as 7/2$^-$ for $^{39,41}$Ar, 5/2$^-$ for 
$^{43}$Ar~\cite{43Ar_BLAUM2008} and then again 7/2$^-$ for $^{45}$Ar ~\cite{F_Lu_2013}.

The $N=26$ $^{44}$Ar is proposed to be deformed with a prolate ground state associated with a 
high $B(E2)$ value for the 2$_1^+$ to 0$_1^+$ transition from 
Coulomb excitation study~\cite{44Ar_coulomb,S_Ar_deformation_Scheit}.  The B(E2) value 
for the same transition in $^{46}$Ar~\cite{Gade_46Ar_2003,46Ar_Calinescu} on the other hand is 
found to be lower supporting the smaller charge 
radius for the $^{46}$Ar~\cite{43Ar_BLAUM2008} compared to $^{44}$Ar. However, the latest 
measurement of the $B(E2)$ values from the lifetime measurement of 
2$_1^+$ state for $^{46}$Ar and $^{44}$Ar by Mengoni {\it et al.}~\cite{44-46Ar_Mengoni_2010}, 
reports almost double $B(E2)$ values for the  $^{46}$Ar than $^{44}$Ar which is unexpected 
if the of $N=28$ shell gap persists for $^{46}$Ar. Different calculations also show
discrepancy between the $B(E2)$ values of these two nuclei which remains to be resolved. 
Between the deformed $^{44}$Ar 
and near spherical $^{46}$Ar, the structure of the intermediate $^{45}$Ar is therefore 
of special interest.

$\beta$-decay is  an excellent experimental tool to study the excited states of 
neutron rich nuclei near $N=28$. The large $Q_{\beta^-}$ value ensures that
a large number of excited states, both bound and unbound, are populated. 
In this region of the chart of nuclides, the
differences in ground-state spins and parities between the $\beta$-decay parent and 
daughter limit the possibility of direct feeding to the ground state. This is due 
to the protons filling the 1$s_{1/2}$ or 0$d_{3/2}$ subshells,
whereas the neutrons are filling the 0$f_{7/2}$ subshell. For odd-$A$ Cl
isotopes like $^{45}$Cl in this work, a positive-parity ground state is expected, 
which will decay to the positive-parity excited states of odd-$A$ daughter Ar 
via allowed Gamow-Teller (GT)
transitions and not feed the negative parity ground state directly.
The positive-parity states in the even-odd daughter arise from promoting a proton 
or a neutron across the $Z=20$ or $N=20$ shell gap and these $1p1h$ states are expected at 
relatively high energies. For $^{45}$Cl, the expected 3/2$^+$ ground state will 
$\beta$-decay to positive-parity (1/2$^+$, 3/2$^+$ or 5/2$^+$) or negative-parity
(1/2$^-$, 3/2$^-$ or 5/2$^-$) states by allowed or first forbidden $\beta$-decays, 
respectively. With the large $Q_{\beta^-}$ value, the $\beta$-decay 
can also populate states above the neutron separation energy (S$_n$) in 
$^{45}$Ar and therefore opens up 
the possibility of studying the excited states in the $\beta 1n$ daughter $^{44}$Ar.
The investigation of excited states in $^{44,45}$Ar is the focus of this study 
along with the ground state spin and parity determination for the parent $^{45}$Cl.

\section{\label{sec:exp}Experimental Setup}

The experiment was carried  out  at  the National  Superconducting  Cyclotron Laboratory
(NSCL) \cite{brad} at Michigan State University to investigate the $\beta^-$ decay of
exotic $^{45}$Cl. A 140-MeV/u $^{48}$Ca primary beam, was  fragmented  using a thick 
Be target at  the  target position  of  the fragment separator, A1900 \cite{A1900}, 
to produce the  nuclei of interest.  
A wedge-shaped Al degrader, which increases the energy dispersion for different 
fragments, was placed at the intermediate dispersive image of the A1900 separator to 
provide a cleaner particle identification of the cocktail beam.

\begin{figure}
	\includegraphics[width=\columnwidth]{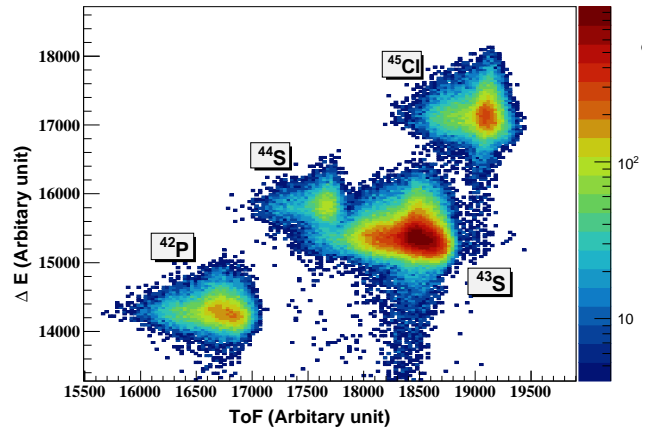}
	\caption{\label{fig:PID_45Clgate}
	 The two dimensional plot of partial energy deposition in the upstream PIN detector 
  ($\Delta E$) and the Time of Flight (ToF) measurement with respect to the focal plane 
  scintillator detector used for particle identification of  the nuclei of interest 
  in the present work. 
	}
\end{figure}

After passing through the wedge shaped Al degrader, the selected isotopes were transported to the 
Beta Counting System (BCS) \cite{prisci}. The BCS is equipped with a $\approx$ 1mm thick 
pixelated (40 strips x 40 strips) Double-Sided Silicon Strip Detector (DSSD) at the center. 
An Al degrader 
upstream reduced the energy of the fragments to ensure that the implants stopped at the 
middle of the DSSD. The DSSD was followed by a Single-Sided Silicon Strip Detector (SSSD) 
which served as a veto detector. This veto detector was used to counter a large flux 
of light particles which were transmitted through the DSSD detector for the particular 
A1900 settings used for this experiment which can impair implant-$\beta$ correlations. 
Dual-gain pre-amplifiers were used for the DSSD to record the time and  position  of implants 
(GeV energy depositions), as well as subsequent decays (keV to MeV energy depositions). 
The implant rate was kept below 200/s to maximize the efficiency of 
correlating the implanted ion with the decay 
 products. 
 
 Two Si PIN detectors, placed upstream of the DSSD, provided the partial energy 
 loss information of the fragments. Along with the scintillator at the intermediate dispersive 
 image of the A1900 these PIN detectors provide the time of flight 
 information used to generate particle identification plots (PID) of the incoming implants as
 shown in Fig.~\ref{fig:PID_45Clgate}. 
 The DSSD and SSSD detector 
 combination was surrounded by 16 Clover detectors to detect the $\beta$-delayed $\gamma$ 
 rays with an efficiency of about $5\%$ at 1 MeV. The efficiency of the array was
 measured with the SRM~\cite{SRM} and $^{56}$Co sources placed outside of 
 the DSSD and then corrected for the dimensions of the DSSD with GEANT4 simulations. 
 The time-stamped data were collected using the NSCL digital data acquisition system 
 \cite{prokop}. The timing and spatial correlations from each channel corresponding 
 to the different detectors were used to ensure the correlation between the implanted 
 fragments in the DSSD and the corresponding $\beta$-decay event.

\section{\label{sec:results}Experimental Results}

Fig. \ref{fig:PID_45Clgate} 
shows the clear separation of the different isotopes produced in the present experimental 
investigation. Selected $^{45}$Cl implants from the cocktail beam were correlated 
with the emitted $\beta$ particles to obtain half-life. Further, coincidence with delayed 
$\gamma$ transitions with the correlated implant-decay events allowed us to study the excited 
states of $^{45}$Ar and $^{44}$Ar produced via $\beta$ and $\beta$-1n decay respectively.
The $\beta$-decay of other neutron rich P and S nuclei seen in Fig.~\ref{fig:PID_45Clgate} 
have been reported and discussed in our previous publication~\cite{Vandana_P_S_decay2022PRC}.

\begin{figure}
	\includegraphics[width=\columnwidth]{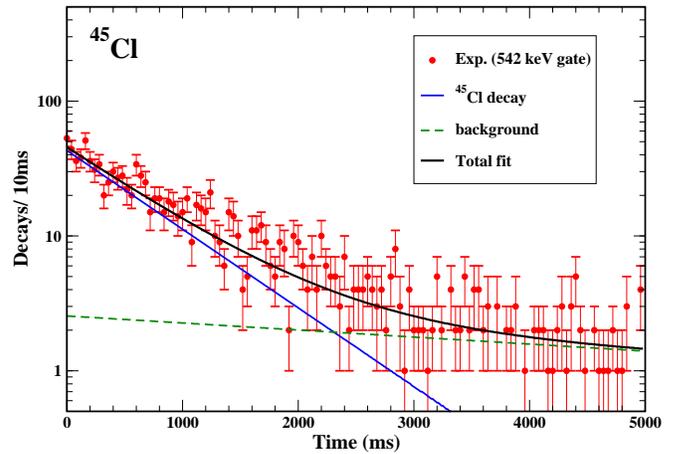}
	\caption{\label{fig:45Cl_deacy}
	The time difference between the $^{45}$Cl implants and correlated $\beta$ decay events 
 gated by the 542-keV ground state transition in $^{45}$Ar. The experimental data is fitted 
 with exponential decay functions to account for $^{45}$Cl decay 
 with suitable background. The half-life is found to be $T_{1/2}$= 513(36) ms. This number is 
larger than the previous measurement of 400(43) ms \cite{Decay_S_Cl_sorlin} though within 2$\sigma$.
	}
\end{figure}

\subsection{\label{beta}$\beta$-decay of $^{45}$Cl}

The time differences between $^{45}$Cl implantations and $\beta$
particles detected in the same or one of the adjacent eight
pixels of the DSSD, in coincidence with the strongest
ground-state $\gamma$ transition (542 keV) in $^{45}$Ar, were histogrammed to generate
a decay curve. The half-life of
$^{45}$Cl was extracted from this decay curve, as shown in
Fig.~\ref{fig:45Cl_deacy}. A fit using a simple exponential decay function of
$^{45}$Cl and a background component accounting for other
long-lived activity gives a half-life of 513(36) ms. The
previous half life measurement for $^{45}$Cl with a value of
400(43) ms is taken from the work of Sorlin {\it et. al.}~\cite{Decay_S_Cl_sorlin}. In
that work the half-lives were deduced from constructing
a time histogram of the  $\beta$-n coincidences detected after the identification
of the corresponding parent nucleus. For $^{45}$Cl 
only 880 events were detected and the authors discuss a possible 
mixing with another implant with shorter half life. 
The measured half life of 513 (36) ms obtained in this work 
is consistent with the shell model calculations using the FSU 
interaction which give a value of 500 ms without including the 
First Forbidden beta transitions.

\begin{figure}
	\includegraphics[width=1.0\columnwidth,angle=0]{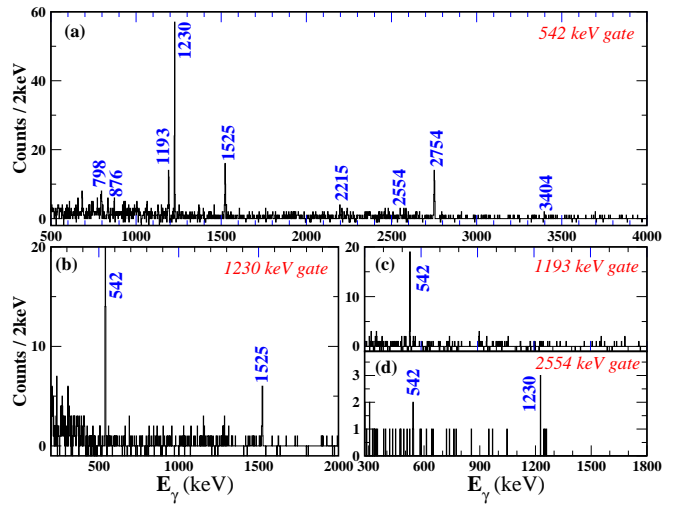}
	\caption{\label{fig:45Ar_coincidence}
	(a)-(d) Coincidences observed between the $\gamma$ transitions in $^{45}$Ar used to 
 establish the level scheme.
	}
\end{figure}

\begin{figure}
	\includegraphics[width=\columnwidth]{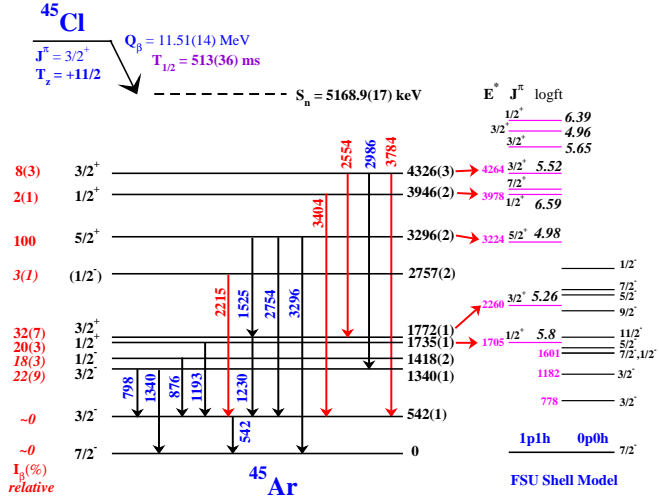}
	\caption{\label{fig:45Ar_levelscheme}
	Partial level scheme of $^{45}$Ar following $\beta^-$-decay of $^{45}$Cl with 
	a $T_{1/2}$ = 513(36) ms and $Q_{\beta^-}$ = 11.51(14) MeV. The transitions 
	marked in black were known before whereas red indicates the new transition (2554-keV) as well
 as prior tentative transitions that we were able to confirm. 
	The branching (relative to 5/2$^+$) is also shown for the excited states. 
        The shell model calculation using the FSU 
 interaction~\cite{Rebeka_SM} predicted the 0p0h and 1p1h excited states and are 
 shown along with the experimental levels.  
	}
\end{figure}

The $\gamma$-$\gamma$ coincidences between the transitions observed in $^{45}$Ar 
are shown in Fig.~\ref{fig:45Ar_coincidence}(a-d) and the corresponding 
level scheme in Fig.~\ref{fig:45Ar_levelscheme}. 
Figure~\ref{fig:45Ar_coincidence}(a)
displays the 542-keV gate which is the strongest transition in $^{45}$Ar and one can see 
all the known transitions: 798, 876, 1193, 1230, 1525, and 2754 keV 
along with the newly placed 2554~keV. The 2554-keV transition is proposed to decay
from the highest observed level, at 4326~keV to the 1772-keV level based on the coincidences
shown in Fig.~\ref{fig:45Ar_coincidence} (d). 
The coincidence gate of 1230~keV shows the 542-keV and 1525-keV transitions only, which 
belong to the same cascade [Fig.~\ref{fig:45Ar_coincidence}(b)]. The absence of any 
transition other than 542~keV in the coincidence gate of 1193~keV confirms that no 
transition is feeding the 1735-keV level [Fig.~\ref{fig:45Ar_coincidence}(c)]. 
In Ref.~\cite{Mrazek_44-45_Ar} two tentative 
$\gamma$ transitions at 3408 and 2215~keV were reported to decay from the proposed 
3950- and 2757-keV levels, respectively. The existence of the 3950-keV (3946~keV in present study)
level was established prior from the neutron transfer study \cite{F_Lu_2013}.  
The present work confirms the presence of 2215- and 3404-keV transitions in the 
coincidence gate of 
542~keV [Fig.~\ref{fig:45Ar_coincidence}(a)].

The level scheme in Fig.~\ref{fig:45Ar_levelscheme} shows the relative 
$\beta$ branching of the levels considering
the absolute efficiencies of the detected $\gamma$-rays. The assignment of the 
spin-parities of the observed levels of $^{45}$Ar is guided by 
predictions from shell-model calculations (discussed later in detail) as well as
from the allowed $\beta$ transition rates from the parent nucleus ($^{45}$Cl). 
The $\beta$ decay should primarily
populate the $1p1h$ positive parity states and hence the log$ft$ values from the SM 
calculations are noted only for the positive parity states. 
Table~\ref{tab:45Ar-44Ar_details} gives the level energies and 
$\gamma$-rays decaying from them, possible $J^\pi$ values, and relative intensities
for both $^{45}$Ar and $^{44}$Ar excited states.

\begin{figure}
	\includegraphics[width=1.0\columnwidth,angle=0]{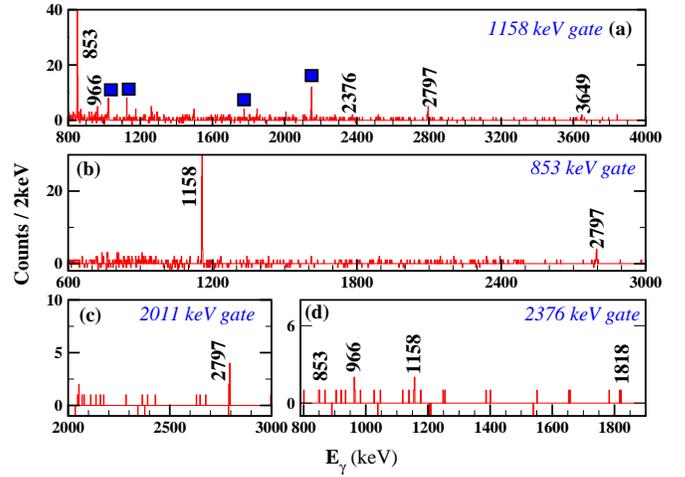}
	\caption{\label{fig:44Ar_coincidences}
    (a)-(d) Coincidences observed between the $\gamma$ transitions in $^{44}$Ar.
    The 1158~keV transition coincidence gate shows contamination from $^{44}$Ca which are 
    marked with blue solid boxes.
	}
\end{figure}

\subsection{\label{betan}$\beta1n$-decay of $^{45}$Cl}

The large $Q_{\beta^-}$ = 11.51 MeV of $^{45}$Cl, along with a low (5.169-MeV) neutron 
separation energy (S$_n$) 
of $^{45}$Ar, leads to a significant beta-delayed neutron branch populating excited
states in $^{44}$Ar. The spin-parities of the states populated in the 
$\beta$-delayed daughter largely depend on the spin-parity of excited state in $^{45}$Ar .

The $\gamma$ rays in coincidence with the 1158-keV ground-state transition 
(2$^+$ to 0$^+$) in  $^{44}$Ar are shown in Fig.~\ref{fig:44Ar_coincidences}(a). 
Here it is worth mentioning that a closeby $\gamma$ transition (1157~keV) exists in 
$^{44}$Ca (2$^+$ to 0$^+$ transition) leading to some spurious coincidences 
marked with solid blue boxes in Fig.~\ref{fig:44Ar_coincidences}(a). 
The strongest transition seen is 853 keV, which is the decay from the 
second 2$^+$, 2011~keV level to the first excited 2$^+$ (1158-KeV) level. This second 2$^+$ state 
at 2011~keV has been confirmed in an earlier deep inelastic reaction study by 
Wan {\it et al.}~\cite{Wan_44Ar_deep}.
The coincidence spectrum also shows the weaker 966-, 2376-, 2797- and 3649~keV transitions
which were already reported in Ref.~\cite{Mrazek_44-45_Ar}. 
The presence and absence of 3649~keV transition in 1158 and 853 gate respectively, 
fixes the placement of this transition from the 4808~keV to the 1158~keV level. 
The highest observed excited level 5354~keV from the present work is seen to be decaying
only to 2978~keV level via the 2376~keV transition.

\begin{figure}
	\includegraphics[width=\columnwidth]{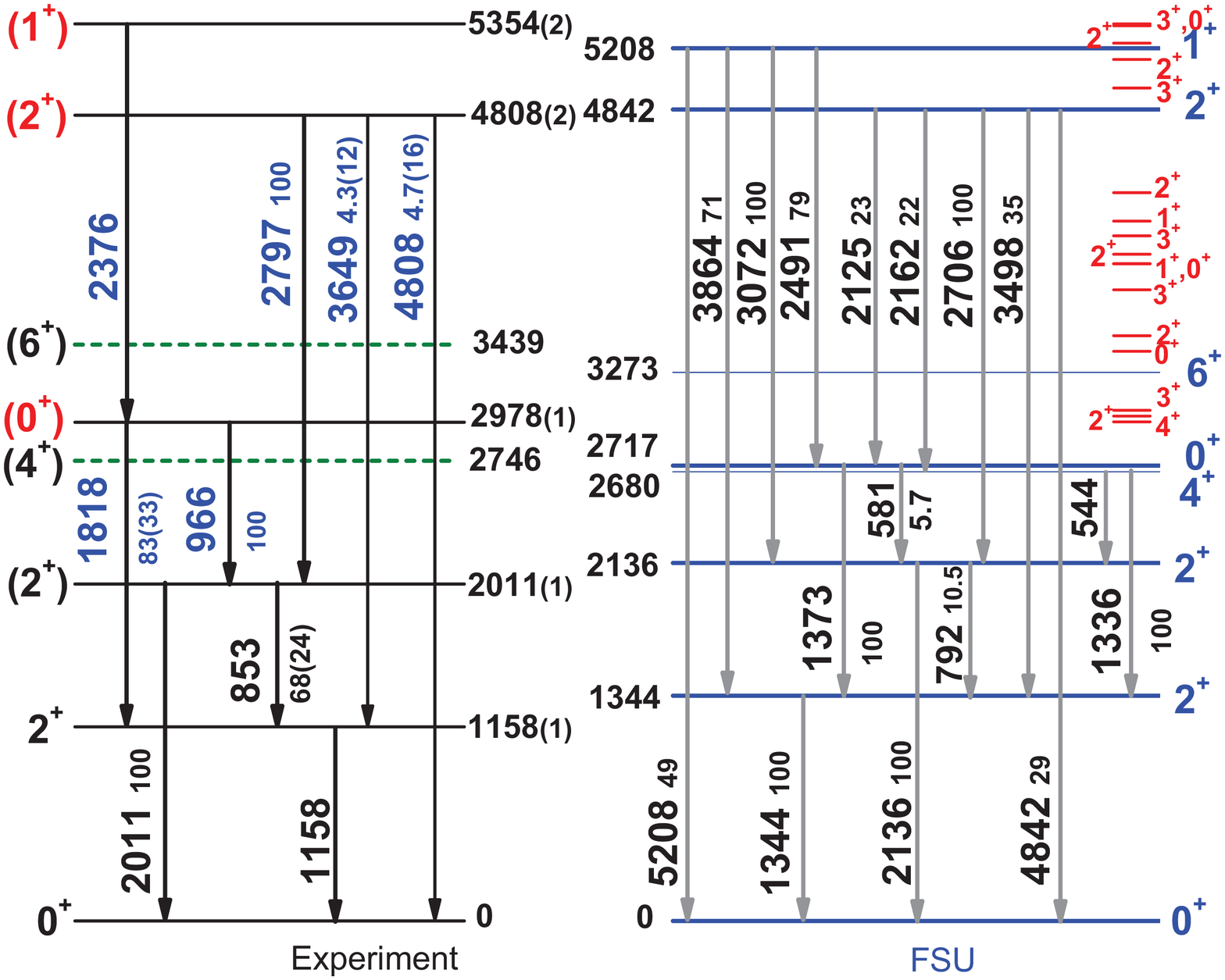}
	\caption{\label{fig:44Ar_ls_SM}
 Partial level scheme of $^{44}$Ar following $\beta 1n$-decay of $^{45}$Cl 
 ($Q_{\beta^- n}$ = 6.34(14)~MeV) is shown in the left panel. 
 The relative intensity of the deexciting transitions 
are also reported along with the associated errors. Two levels are shown by green 
dashed lines which were reported previously but not seen in the present work. 
The transitions which are seen for the first time from $\beta 1n$ work are marked in blue. 
The spins of the 
levels that are suggested for the first time in the present work are marked with red 
(see section B of the discussion).
 Predictions of shell-model (SM) calculations using the FSU 
 interaction~\cite{Rebeka_SM} are shown in the right panel. The possible 
 $\gamma$-transitions and their relative branching ratios were also calculated and are noted. 
 The excited states, predicted by the SM but not observed in the experiment, are shown 
 at the right of the SM level scheme in red.
	}
\end{figure}

The level scheme, formed from the present work is shown in Fig.~\ref{fig:44Ar_ls_SM} 
(left panel). Along with the experimental levels, the corresponding SM predicted levels 
(right panel) are also shown. All the transitions, apart
from the 853-, 1158- and 2011-keV transitions are observed for the first time 
in the $\beta 1n$ decay of $^{45}$Cl and are marked in blue in the experimental level
scheme shown in Fig.\ref{fig:44Ar_ls_SM} (left panel).
Two levels at 2746~keV and 3439~keV, shown in the experimental level scheme by
green dashed line, were not populated in the $\beta$n channel but are shown for 
comparison with 
the shell model calculations and were seen in the previous study by Fornal {\it et. 
al.}~\cite{44Ar_deep_Fornal}.

\section{\label{sec:discuss}Discussion}

\subsection{\label{45Ar_dicsussion}$^{45}$Ar}

The ground state spin parity of $^{45}$Cl is expected to be 3/2$^+$ due to an odd proton in 
$d_{3/2}$ orbital and a full $\nu$$f_{7/2}$ orbital in a simple picture as illustrated in 
Fig.~\ref{fig:45Cl_decay_pattern}. However 
with the $d_{3/2}$ and $s_{1/2}$ orbitals being nearly degenerate, a $1/2^+$ assignment  
cannot be ruled out. Below, based on the $\gamma$ decay characteristics and the shell-model 
calculations, we suggest a $3/2^+$ spin-parity. 
The shell-model calculations presented 
in this work were performed using the shell model code CoSMo~\cite{volya} using the 
FSU interaction. The FSU 
interaction is a data-driven shell-model interaction aimed to explain cross-shell 
excitations between the $sd$ and $fp$ shell and also $p$ and $sd$ shells for 
neutron-rich $sd$-shell nuclei~\cite{Rebeka_SM}. 
The predictions of the FSU interaction have found great 
success in explaining many experimental observations~\cite{Gade2022,Brown2022}.   
For the calculations quoted here, when the excitations are confined to 
the major shells ({\it e.g.} $sd$ and $fp$) they are referred to as $0p0h$ excitations, whereas 
$1p1h$ calculations involve movement of a single nucleon between the major shells.

\subsubsection{\label{45Ar_positive}Positive-parity states}

From the selection rules of allowed Gamow-Teller (GT) decay, 
states with positive parity should have the highest branching in the daughter nucleus. 
In a more nuanced picture, the $\beta^-$ decay will likely involve the conversion of a
$d_{3/2,5/2}$ neutron into  a $d_{3/2}$ proton as $^{45}$Ar has a vacancy 
in the $\pi d_{3/2}$ orbital. The other possibility is for the $s_{1/2}$ neutron to 
transform into an $s_{1/2}$ proton as indicated  in 
Fig.~\ref{fig:45Cl_decay_pattern}. These will create neutron hole states in $^{45}$Ar 
corresponding to $1p1h$ states in the shell-model calculations. 
We propose the states at 1735, 1772, 3296, 3946, and 4326 keV 
with the highest branching to be populated via allowed $GT$ transitions 
and hence have positive parity. In a recent transfer study using the reaction 
$^1$H($^{46}\mathrm {Ar},d)^{45}\mathrm{Ar}$ \cite{F_Lu_2013} the same states were proposed as 
neutron hole states. The observed states at 1735 and 1772 keV  in 
Ref.~\cite{F_Lu_2013} were found to have large spectroscopic factors for $\nu d_{3/2}$ and 
 $\nu s_{1/2}$ hole states but, due to limited experimental energy resolution, 
 they were not able to resolve the two levels. 
 In this work, with the excellent resolution of the high purity Germanium detectors, we 
 are able to assign accurate energies to these two. 
  The shell-model calculations presented, found the 3/2$^+$ state at a higher energy than the 
 1/2$^+$ state and as a result, we assign the 1735- and 1772-keV levels as 1/2$^+$ and 3/2$^+$, respectively.
 This is further confirmed by the branching ratio of the decays of the 3296-keV state as 
 discussed ahead. 
 
The 3296-keV state has the strongest population in the current $\beta$-decay 
 study. It has decay branches to the 
 $7/2^-$ ground state, the 542-keV first excited $3/2^-$ state and a strong branch to the 
 1772-keV state which we propose to be the $3/2^+_1$ state. We could not identify any transition
 to the 1735-keV state within our observation capability. As the 3296-keV state decays 
 to the ground state, it likely has a $5/2^+$ spin-parity. The vicinity 
 of this state to the $5/2^+$ 3224-keV state predicted in the SM calculations ratifies 
 this assignment and the calculated log$ft$ value of 4.98 justify its strong population in beta decay.
 The predicted decay 
 probabilities of the $5/2^+_1$ states to the $3/2^+_1$ and $1/2^+_1$ states in SM calculations 
 are listed in Table~\ref{tab:45Ar_branchingratio}. 
 The decay to the $3/2^+$ state via a M1 transition is stronger and supports the spin 
 assignments to the doublet of states at $\approx 1.7$~MeV. 
\begin{table}[b]
  \caption{\label{tab:45Ar_branchingratio}
  	SM predictions (using the FSU interaction) for the branching ratio of the $5/2^+_1$ state 
 in $^{45}$Ar. The experimental value of the $\gamma$ ray transition was used in the calculation 
 of the rates. The difference in branching to the $3/2^+$ and $1/2^+$ states is used to confirm 
 the spin parity assignment to the two experimental state at $\approx$1.7 MeV.}
 
		\begin{tabular}{cccc}
			\hline
		$J_i$ $\rightarrow$ $J_f$~~~&~~~ $E_\gamma$ (keV)~~~ &B(M1) ($\mu_N^2$)~~&B(E2)(e$^2$fm$^4$)\\
                     -~~~&~~~ -               ~~~~& rate(1/sec) & rate(1/sec)\\	  
			\hline
			5/2$^+$ $\rightarrow$ 1/2$^+$ & 1525 & - & 104\\
             -                      &   -  & - &  1.05e+12\\
             \hline
             5/2$^+$ $\rightarrow$ 3/2$^+$ & 1525 & 0.35 & 27.3\\
             -                      &   -  & 2.19e+13 &  2.77e+11\\
			\hline		
		\end{tabular}
\end{table}
 
\begin{figure}
	\includegraphics[width=\columnwidth]{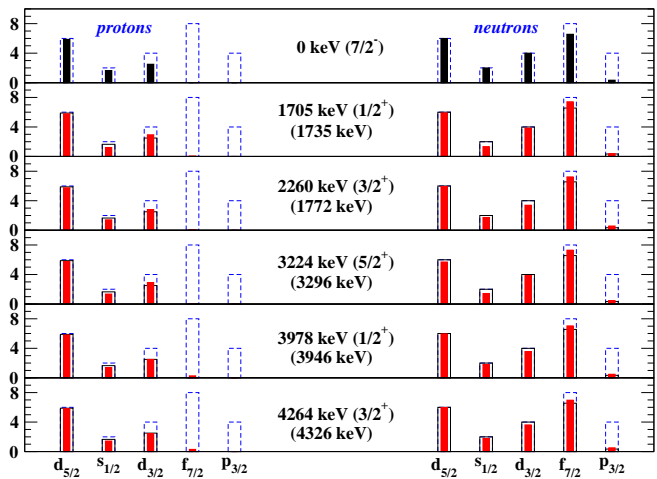}
	\caption{\label{fig:45Ar_occ}
 The occupation number of different orbitals (proton and neutron) for the positive parity 
 excited states calculated from shell-model using the FSU interaction~\cite{Rebeka_SM} for $^{45}$Ar are 
 shown. The energies of the experimental states corresponding to each proposed SM states in the 
 present work are indicated in parentheses. The blue dashed columns are the maximum occupancy for an orbital ($2j+1$).
 The solid black columns are the occupancy of the 7/2$^-$ ground state and the red 
 columns are the occupancies of the excited states of $^{45}$Ar.
 }
 \end{figure}

 The occupation numbers for the ground state and the excited positive-parity states in $^{45}$Ar 
  from SM calculations are shown in Fig.~\ref{fig:45Ar_occ}. 
  The occupation numbers for the two states at $\approx 1.7$~MeV clearly 
 establish them as $\nu$$s_{1/2}$ and $\nu$$d_{3/2}$ hole states, respectively. 
  The occupancy of the 3296-keV state from the shell-model calculation shows
 contribution from the $\nu$$d_{5/2}$ hole along with $s_{1/2}$.
 This level was also observed via the $^1$H($^{46}$Ar,$^{45}$Ar)$^{2}$H reaction by Lu $\ {et. 
 al.}$~\cite{F_Lu_2013} though no spin was assigned to that state. 
 The assignment of $5/2^+$ for 3296~keV from the present work, as discussed above, 
 is consistent with the observation of a small peak at 3.29~MeV~\cite{F_Lu_2013} in the 
 deuteron spectra. 
 Another state at 3.95~MeV, described to have $\ell =0$ parentage in Ref.~\cite{F_Lu_2013},  
 is also observed in the present work. The shell-model calculation supports the presence of 
 the second 1/2$^+$ state at 3978~keV with considerable contribution from both neutron 
 and proton $s_{1/2}$ orbitals. Therefore, the state at 3946~keV is assigned a 1/2$^+$ spin-parity.
  The shell model further predicts a closeby 3/2$^+$ at 4264~keV 
 (see Fig.~\ref{fig:45Ar_levelscheme}), which we have associated with the 4326-keV state. 
 The higher energy states (at 3978~keV and 4264~keV) show a rise in occupation
 number for protons in the $f_{7/2}$ orbital with respect to the ground-state configuration 
 while the occupation number for $\pi$$d_{3/2}$ remains the same. It is accompanied by a 
 slight drop in the occupancy for the $\nu$$f_{7/2}$ orbital. These states could 
have contribution from conversion of an $f_{7/2}$ neutron to a $f_{7/2}$ proton in 
the beta  decay.

\subsubsection{\label{45Cl_gs} Ground state of parent $^{45}$Cl}
 The strong population of the $5/2^+$ at 3296 keV in the $\beta$ decay leads to the 
 determination of the ground state spin and parity of the parent to  be $3/2^+$. 
 The ground-state spin-parity of the neutron-rich odd-mass Cl isotopes are in the spotlight 
 due to the degeneracy of $\pi$$s_{1/2}$ and $\pi$$d_{3/2}$ orbitals as one moves from 
 the $N=20$ $^{37}$Cl to
 the neutron-rich $^{45}$Cl ($N$=28). Gade {\it et al.}~\cite{Gade_K_Cl_P_ProtonOrbital} 
 systematically showed the reduction of the $E$(1/2$^+$)-$E$(3/2$^+$) gap as a function
 of neutron-proton asymmetry for all the odd-mass K, Cl and P isotopes. 
 Though the ground-state spins have
 been experimentally verified for $^{41}$Cl and $^{43}$Cl as $1/2^+$ with a very closeby
 $3/2^+$, the tentatively assigned $1/2^+$ ground state of $^{45}$Cl based on SM 
 calculations in Ref.~\cite{Gade_K_Cl_P_ProtonOrbital} had not been confirmed 
 experimentally. Two closely spaced energy states (127 keV apart) are predicted 
 as candidates for the $1/2^+$ and $3/2^+$ generated from 
 proton holes in $s_{1/2}$ and $d_{3/2}$ respectively. The present work reports the highest
 $\beta$-feeding to the $5/2^+$ state in $^{45}$Ar from the gs of $^{45}$Cl. This is 
 possible only from a $3/2^+$ ground state and can be considered as the first experimental 
 support for the assignment of $3/2^+$ spin to the ground state over $1/2^+$ for $^{45}$Cl 
 indicating a return of normal filling of orbitals in odd-A Cl isotopes. $\beta$-decay 
 of $^{45}$S into $^{45}$Cl can shed further light on the ground state spin/parity.

\subsubsection{\label{45Ar_negative}Negative parity states}
 The lower states in $^{45}$Ar are supposed to have negative parity arising from the
 excitations of the odd neutron(s) within the $fp$ shell. Their population in $\beta$-decay 
 is either due to 
 feeding from the high-lying states or could also arise from First Forbidden (FF) transitions.

The excited states at 1340 and 1418 keV are found to be consistent with the
previous $\beta$-decay study by Mrazek {\it et al.}~\cite{Mrazek_44-45_Ar}.
The state at 1.34~MeV was tentatively reported by Lu {\it et al.}~\cite{F_Lu_2013}
in a $^1$H($^{46}$Ar, d)$^{45}$Ar study though the $^{45}$Ar(d, p)$^{45}$Ar reaction 
Gaudefroy {\it et. al.} ~\cite{Gaudefroy_2005,45Ar_dp_2008} couldn't identify this state.
Gaudefroy {\it et. al.} on the other hand, reported a 1420(60)-keV excited state with $3/2^-$ spin-parity and 
described it as a member of the multiplets of the $\pi$(2$^+$)$\otimes$$\nu$$f_{7/2}$ configuration 
with one proton hole in $s_{1/2}$ and another in $d_{3/2}$.
Because of the mixed nature, both these states were weakly populated in either of the 
transfer-reaction study~\cite{45Ar_dp_2008,F_Lu_2013}. 
The present shell-model calculation predicts the second 
excited state to be $3/2^-$ with a closeby $1/2^-$ state (see Fig.~\ref{fig:45Ar_levelscheme}).
The 1340-keV level has a direct $\gamma$-decay branch to 
the 7/2$^-$ ground state favoring a 3/2$^-$ spin assignment over 1/2$^-$ spin whereas 
the 1418-keV state decays to only the 542-keV $3/2^-$ state
and not to the ground $7/2^-$ state.
Therefore, it is proposed that the second excited state at 1340 keV has a 3/2$^-$ spin-parity 
while the 1418-keV level is a 1/2$^-$ consistent with the $\ell = 1$ assignment 
of Ref.~\cite{Gaudefroy_2005} as well as present shell model predictions with a
composite configuration of $\pi$($d_{3/2}$ $\otimes$ $s_{1/2}$) $\otimes$ $\nu$$p_{3/2}$.

\begin{figure}
	\includegraphics[width=\columnwidth]{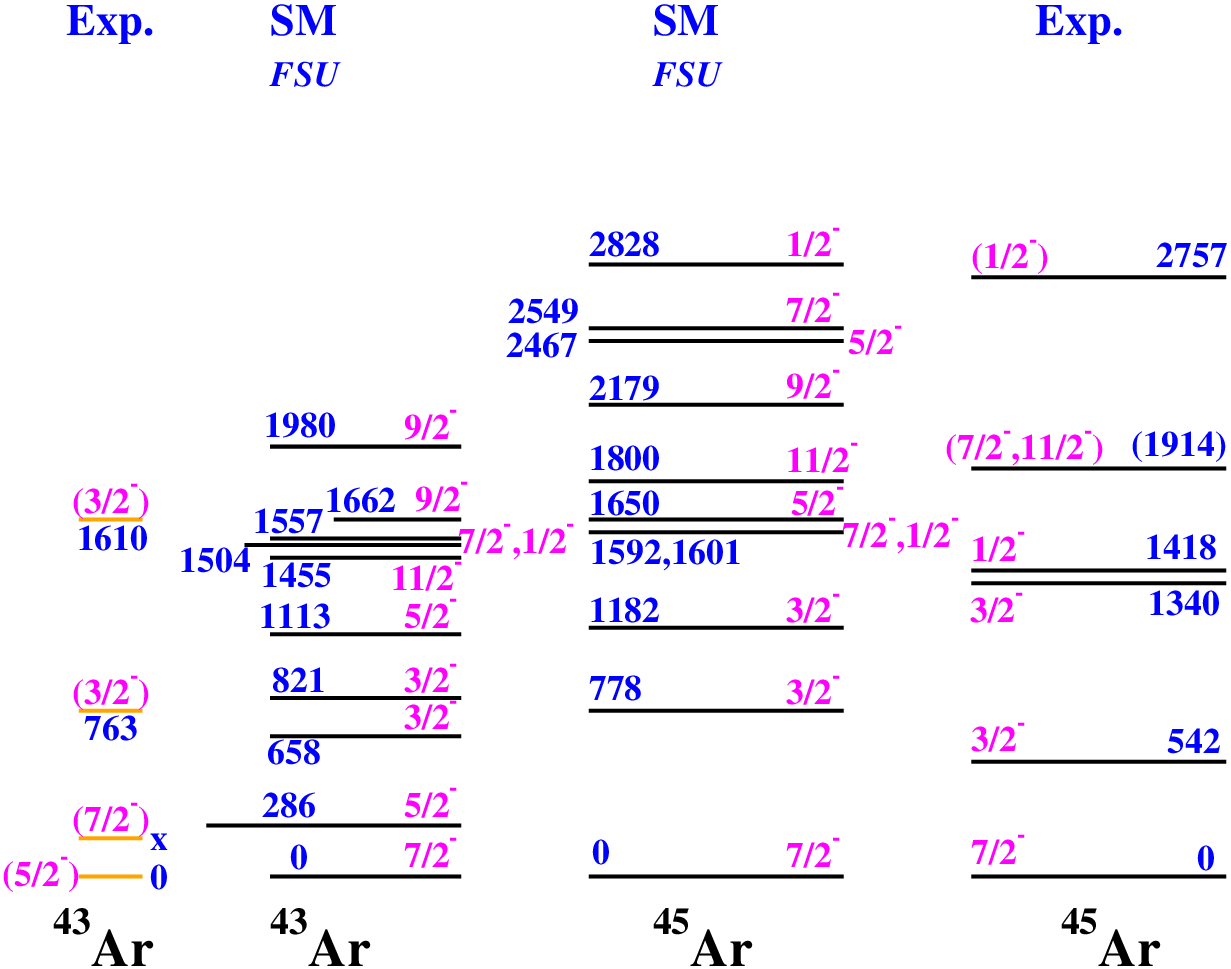}
	\caption{\label{fig:45Ar_sys}
 The low-energy negative-parity states are displayed for $^{45}$Ar and $^{43}$Ar~\cite{Ar_Szilner_transfer,nndc} 
 for comparison both from 
 experiment and shell-model calculations using the FSU interaction~\cite{Rebeka_SM}. The experimental energy of $7/2^-$
state for $^{43}$Ar is yet unknown.}
\end{figure}

The negative-parity energy levels in $^{45}$Ar (observed here and from prior studies) and 
$^{43}$Ar are further compared with the shell-model calculations
in Fig.~\ref{fig:45Ar_sys} to understand the evolution of the $N=28$ shell gap. 
Relatively less information is available for the negative parity states in $^{43}$Ar, as is clear 
in the figure. 
Unlike $^{45}$Ar, a ground-state doublet of $5/2^-$ and $7/2^-$ spins is
predicted in $^{43}$Ar and can be considered as members of the 
multiplets arising from the configuration $\pi d_{3/2}^{-2} \nu f_{7/2}^{-3}$.
There is only tentative experimental evidence of this doublet in $^{43}$Ar with 
5/2$^-$ proposed to be the ground state~\cite{Ar_Szilner_transfer,43ArProtonScat}.
Further the $\beta$-decay of $^{43}$Cl with a 3/2$^+$ gs
\cite{43Ar_Winger2006} shows a large branch to the ground state of $^{43}$Ar through First 
Forbidden(FF) decay which  also favors the $5/2^-$ assignment.  
Hence, though the SM calculations using the 
FSU interaction correctly predict very closely spaced 7/2$^-$(gs) and 5/2$^-$ state 
(286~keV in Fig.~\ref{fig:45Ar_sys}) for $^{43}$Ar, the 5/2$^-$ is more probable 
for the ground state. For $^{45}$Ar on the other hand, both the 
SM calculations and the experimental observations do not support a
close 5/2$^-$-7/2$^-$ ground-state multiplet, a signature of the proximity  to the 
$N=28$ shell closure.

 \subsection{\label{44Ar_dicsussion}$^{44}$Ar}

The states populated in $^{44}$Ar follow the neutron emission from the $1p1h$ 
positive parity states with spins of $1/2$, $3/2$ or $5/2$ populated in $^{45}$Ar.
The first excited state of even-even $^{44}$Ar is at 1158~keV  with a $J^\pi$ of 
2$^+$ known from earlier Coulomb-excitation, in-beam $\gamma$-spectroscopy and deep 
inelastic studies~\cite{S_Ar_deformation_Scheit,44Ar_coulomb,Wan_44Ar_deep,44Ar_deep_Fornal} and 
is described as a deformed state while the second excited one is also a 2$^+$ state at 2011 keV.
The spins of the other excited levels observed at 2978, 4808 and 5354~keV are proposed to be
0$^+$ to 4$^+$ in NNDC~\cite{nndc44Ar}. We have tried to make more specific spin 
assignments to these by comparing with the predictions of shell-model calculation in the
$0p0h$ valence space. The experimental states, though, are likely to have some contribution 
from $2p2h$ configurations as the neutron is likely being emitted 
from the $sd$ shell. Our calculations currently cannot accommodate this.

\begin{table}
  \caption{\label{tab:45Ar-44Ar_details}
      $\gamma$-ray energies along with the corresponding initial levels and initial and final spins for $^{45}$Ar and $^{44}$Ar observed in the present
work are presented. For $^{45}$Ar the intensities of the $\gamma$ rays are normalized with respect to the strongest 542-keV $\gamma$ ray. For $^{44}$Ar the branching from each level is shown with each branch normalized to
 the strongest one from that level.}

		\begin{tabular}{cccc}
			\hline
   $^{45}$Ar\\
   \hline
\hline
			  $E_i~$~&~~$J_i$ $\rightarrow$ $J_f$~~~&~ $E_\gamma$ ~ & I$_{rel}$\\
     (keV)~&~~~~~&~ (keV)~ & \\
			\hline
			
			542(1)~~&~3/2$^-$ $\rightarrow$ 7/2$^-$ & 542(1) & 100(10)\\
            
             \hline
			1340(1)~~&~3/2$^-$ $\rightarrow$ 7/2$^-$ & 1340(1) & 11.4(13)\\
            ~~&~3/2$^-$ $\rightarrow$ 3/2$^-$ & 798(2) & 5.7(7)\\
             \hline	
			1418(2)~~&~1/2$^-$ $\rightarrow$ 3/2$^-$ & 876(1) & 11.4(12)\\
             \hline	
			1735(1)~~&~1/2$^+$ $\rightarrow$ 3/2$^-$ & 1193(1) & 13.0(15)\\
            \hline	
            1772(1)~~&~3/2$^+$ $\rightarrow$ 3/2$^-$ & 1230(1) & 37.0(39)\\
             \hline	
             2757(2)~~&~(1/2$^-$) $\rightarrow$ 3/2$^-$ & 2215(2)& 2.1(4)\\
             \hline
			3296(2)~~&~5/2$^+$ $\rightarrow$ 3/2$^+$ & 1525(1) & 18.0(20)\\
            ~~&~5/2$^+$ $\rightarrow$ 3/2$^-$ & 2754(1) & 29.0(32)\\
            ~~&~5/2$^+$ $\rightarrow$ 7/2$^-$ & 3296(2) & 12.1(16)\\
             \hline
             3946(2)~~&~1/2$^+$ $\rightarrow$ 3/2$^-$ & 3404(2)& 1.3(4)\\
             \hline
             4326(3)~~&~3/2$^+$ $\rightarrow$ 3/2$^+$ & 2554(2) & 2.6(5)\\
            ~~&~3/2$^+$ $\rightarrow$ 3/2$^-$ & 2986(2) & 2.1(4) \\
            ~~&~3/2$^+$ $\rightarrow$ 3/2$^-$ & 3784(3) & 3.7(7)\\
             \hline
   $^{44}$Ar& ~~~~~&~~~~~~~~~& Rel. Branching\\
   \hline
\hline
		
			1158(1)~~&~2$^+$ $\rightarrow$ 0$^+$ & 1158 (1) & 100 \\
            \hline
			2011(1)~~&~(2$^+$) $\rightarrow$ 2$^+$ & 853(1) & 68(24)\\
            ~~&~(2$^+$) $\rightarrow$ 0$^+$ & 2011(1) & 100\\
             \hline
			2978(1)~~&~(0$^+$) $\rightarrow$ (2$^+$) & 966(1) & 100 \\
            ~~&~(0$^+$) $\rightarrow$ 2$^+$ & 1818(1) & 83(33)\\
            \hline
			4808(2)~~&~(2$^+$) $\rightarrow$ (2$^+$) & 2797(2) & 100 \\
            ~~&~(2$^+$) $\rightarrow$ 2$^+$ & 3649(2) & 4.3(12)\\
            ~~&~(2$^+$) $\rightarrow$ 0$^+$ & 4808(2) & 4.7(16)\\
	\hline
			5354(2)~~&~(1$^+$) $\rightarrow$ (0$^+$) & 2376(1) & 100 \\		
   \hline		
		\end{tabular}
\end{table}

The experimental levels (left panel) and SM predicted states (right Panel)
for $^{44}$Ar are shown in Fig.~\ref{fig:44Ar_ls_SM} along with the relative 
intensities of the $\gamma$-transitions from each level. 
The experimental 2978-keV level decays to the first and 
second 2$^+$ states via the 1818- and 966-keV transitions, respectively, 
where the 966-keV decay dominates over the 1818-keV branch.
This 2978-keV level lies close in energy to 4 predicted states (0$^+$, 4$^+$, 2$^+$ and 3$^+$). 
The calculated 4$^+$ (2978~keV) and 2$^+$ (3013~keV) states have higher transition rates to the 
4$^+_1$ (2680~keV) and 0$^+$ (gs) states, respectively, which is not observed in the decay of 
experimental 2978-keV level. 
This leaves the two spins, 3$^+$ (3047~keV) and 0$^+$ (2717~keV), as the most 
probable candidates for this level. 

\begin{figure}
	\includegraphics[width=\columnwidth]{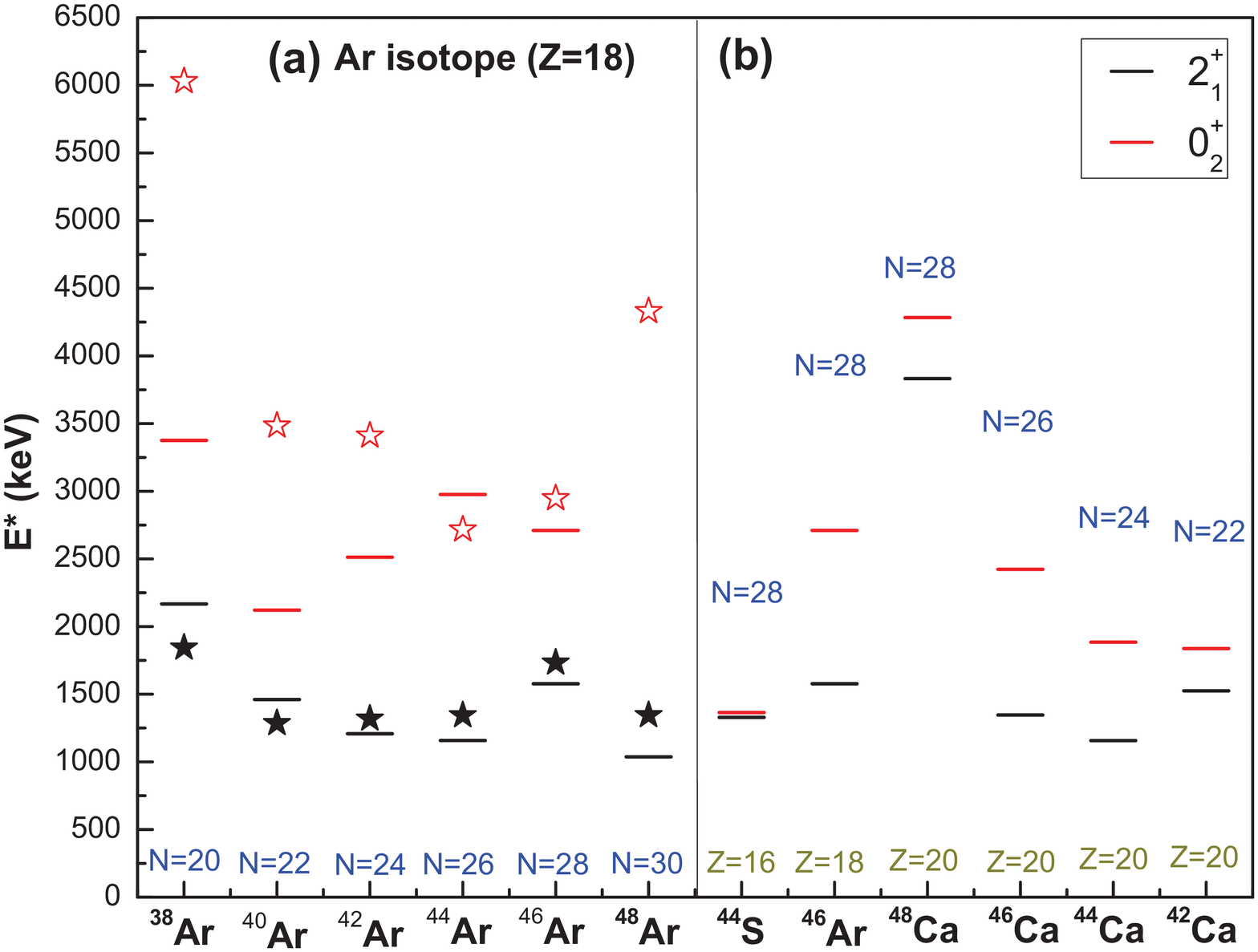}
	\caption{\label{fig:Ar_def}
(a) The experimental low-lying 2$_1^+$ and 0$_2^+$ energies of Ar isotopes as a function of neutron number. 
The shell-model (using the FSU interaction) predicted states are also shown as closed (open)
star for 2$_1^+$ (0$_2^+$) spin. The 0$_2^+$ spin for $^{42}$Ar is adopted from one of the 
possible spins predicted in NNDC~\cite{nndc} for the 2512.5-keV level for comparison purposes. 
(b) Comparison of 2$_1^+$ and 0$_2^+$ energies in Ca, Ar and S nuclei near magic numbers 
$N=28$ and $Z=20$. From the doubly magic $^{48}$Ca (in center), the left (right) isotones 
(isotopes) have two protons (neutrons) less than the previous one, keeping the 
$N=28$ ($Z=20$) magic number.
The experimental values for the nuclei (other than $^{44}$Ar) are taken from 
Refs.~\cite{38Ar_Speidel_PLB,S_Ar_deformation_Scheit,48Ca_Vanhoy1992,Ar_Szilner_transfer,40Ar_Speidel_2008, 48Ar_Sarmi_PRL,44-46Ca_Peng_PRL,nndc}. 
}
\end{figure}

The level at 4808~keV decays to the ground state and 
the excited 2$^+$ (1158- and 2011-keV) states. Therefore, among the prior suggested spin-parity 
assignments, 4$^+$, 3$^+$ or 0$^+$~\cite{nndc44Ar} are not possible for this state. 
Between the remaining 2$^+$ and 1$^+$ spin, the shell model does not predict any 1$^+$ 
state nearby (Fig.~\ref{fig:44Ar_ls_SM}), therefore the spin for this
level is suggested to be 2$^+$ corresponding to the level 4842~keV from the calculation.
Further, the  calculations (Fig.~\ref{fig:44Ar_ls_SM}) predict 
that the 4842-keV level (exp. 4808~keV level) has the most intense decay branch 
to the gs and the excited 2$^+$ states, which matches with the experimental observation.
If the spin-parity of 2978-keV state is assigned as 3$^+$ then the shell-model predicts a strongest decay branch
from 4808-keV to 2978-keV level.
The absence of a decay path from 4808-keV to 2978-keV state encourages us to assign the 2978-keV state 
as the excited $0^+$ over the $3^+$ possibility. The systematics of excited 
0$^+$ in Ar isotopes will be discussed next.

The highest observed level from the present $\beta$n decay work is at 5354~keV 
with a decay only to the newly assigned 0$^+$ 2978-keV level. This 5354-keV level was 
proposed  to decay to the 2011-keV (2$^+_2$) and 1158-keV (2$^+_1$) states via the
3342- and 4195-keV transitions in the earlier $^{44}$Cl beta-decay 
work~\cite{Mrazek_44-45_Ar} but with 3-5 times less intensity than the 2376~keV transition. 
The observation of the strong 2376-keV transition to the 2978-keV 
(0$^+$) state rules out 0$^+$ and 3$^+$ assignments for the 5354~keV level. 
The shell model predicts a 2$^+$ at 5141~keV and a 1$^+$ at 5208, both of which are good candidates but the
2$^+$ is predicted to decay by a strong transition to the 2$^+_2$ state not consistent 
with the experiment. Therefore, the 5354 keV is assigned as 1$^+$, consistent with all 
experimental observations. 
The population of 1$^+$ and 2$^+$ in the delayed neutron decay suggests the population 
of a $3/2^+$ unbound state in $^{45}$Ar which decays by a $\ell = 0$ neutron.

\subsection{\label{Ar isotopes}Even-Even Isotopes near $N=28$}

To understand the evolution of shape away from the $N=28$ shell closure, the excited state 
energies of 
2$_1^+$ and 0$_2^+$ excited states are plotted for even-even Ar isotopes as a function of neutron number in 
Fig.~\ref{fig:Ar_def}(a). With $N=20$, both the 2$_1^+$ and 0$_2^+$ are 
high in energy for $^{38}$Ar, reflecting the large shell gap between the $d_{3/2}$ and $f_{7/2}$ 
orbitals. As we increase the neutron number approaching half occupancy of the $f_{7/2}$ orbital, 
the lowering of the first 2$^+$ state indicates an increasing collectivity, and reduction 
of the $N=28$ shell gap. 
In the Ar isotopic chain, it is interesting to notice that the most collective behavior 
is for the ground state of $^{44}$Ar ($N=26$) signified by the lowest energy of 2$_1^+$.
After that, the increase of the 2$_1^+$ energy points towards the restoration of 
the shell gap between the $\nu$$f_{7/2}$ and $\nu$$p_{3/2}$ orbitals in $^{46}$Ar.
With an additional neutron pair above $N=28$, the 2$^+$ state comes down in energy 
for $^{48}$Ar.

A different trend is seen for the 0$_2^+$ state, which generally represents a 
different shape of the nucleus from the ground state. For $^{40}$Ar, the 0$_2^+$ state is 
described to be part of a super-deformed band in Ref.~\cite{40Ar_Ideguchi_sd_band2010}. 
With increasing neutron number, the energy of this state 
is found to increase, attaining a maximum value for $^{44}$Ar.
After that, the experimental 0$_2^+$ state shows a decreasing trend again for $^{46}$Ar. As can be seen 
from Fig.~\ref{fig:Ar_def}(a), the shell-model calculations with the FSU interaction 
(solid black stars) closely mirror the 2$_1^+$ values for $^{38-48}$Ar. 
For the 0$_2^+$ states (open red star) from $0p0h$ configurations, the shell-model 
predictions show an increasing trend in energy with decreasing neutron number, in 
disagreement with the experiment. 
It may be inferred that the 0$_2^+$ states for $^{38-42}$Ar isotopes
have a contribution from $2p2h$ configuration which is beyond the scope of 
present SM calculations.
It will be interesting to search for the 0$_2^+$ state in $^{48}$Ar, 
which is predicted to be very high (4.3 MeV) in the present SM calculation. 

For further systematic analysis, experimental 2$_1^+$ and 0$_2^+$ states are 
plotted for selected nuclei around $Z=20$, $N=28$ magic shell closures in 
Fig.~\ref{fig:Ar_def}(b). The doubly magic $^{48}$Ca (in the middle) shows a high lying 
2$_1^+$ and excited 0$_2^+$ representing a pronounced $Z=20$, $N=28$ shell gap. 
With two fewer protons, for $^{46}$Ar the 2$_1^+$ and 0$_2^+$ are lower in energy
with a further decrease in the state for $^{44}$Ar [Fig~\ref{fig:Ar_def}(a)] 
indicating the deformation associated with the ground state.
Reducing two more protons, for $^{44}$S the $2^+_1$ drops further and importantly
the spacing between the 0$_2^+$ and 2$_1^+$ levels collapses resulting in low lying 
prolate-spherical shape co-existence in $^{44}$S~\cite{44S_Force2010,44S_shape_warner}. 
In contrast, for the isobar $^{44}$Ar, the $2^+_1$ and 0$_2^+$ show a large separation. 
In Fig.~\ref{fig:Ar_def}(b) the isotopes of Ca away from the $N=28$ shell closure are also 
shown to the right of $^{48}$Ca and one can notice the same trend of reduction 
of the gap between the 2$_1^+$ and 0$_2^+$ states along with lowering of the energy of 
0$_2^+$. 
Therefore, the systematics suggests that if we decrease either the proton or neutron number 
away from doubly magic 
$^{48}$Ca, keeping the other magic number constant, the 0$_2^+$ which broadly represents 
a different shape in excited state moves lower in energy, approaching the possibility of 
shape coexistence. But with both proton and neutron number away from the magic number, 
the nuclei seem to favor one shape at low excitation energy.

\section{\label{conclude}Summary}

The $\beta^-$ decay of $^{45}$Cl is reported here, from an experiment performed at the NSCL 
following the fragmentation of a $^{48}$Ca primary beam.
The half-life ($T_{1/2}$) of $^{45}$Cl is measured to be 513(36) ms, which is longer 
than the prior measurement from GANIL but consistent with shell model calculations using the 
FSU interaction. The level schemes of $^{45}$Ar and $^{44}$Ar are established from the 
observed $\gamma$-$\gamma$ coincidences in the $\beta$ and 
$\beta 1n$ channels, respectively.
Many of the prior tentative placements of transitions in $^{45}$Ar 
have been verified and a new $\gamma$-transition at 2554~keV has been added.  
The experimentally observed levels are compared with SM calculations  
for both $^{44,45}$Ar with excellent agreement. 
From the predicted occupancy of different orbitals and the decay pattern 
of the $\gamma$ transitions from the excited levels, the spin-parity of the levels of 
$^{45}$Ar populated via GT transitions are proposed.
The higher lying positive-parity states of $^{45}$Ar are candidates for
$1p1h$ excitations consistent with their population in prior transfer reactions. 
The maximum feeding to the 5/2$^+$ state in $^{45}$Ar, supported by the 
small log$ft$ value calculated from SM calculations, allowed us to assign a spin parity 
of $3/2^+$ to the ground state of the parent $^{45}$Cl. 
The spin and parity for the levels in $^{44}$Ar are proposed by comparing with 
SM calculations. An excited 0$_2^+$ state is proposed for the first time in $^{44}$Ar
at 2978~keV. The SM calculations reproduce the experimental evolution of 
the 2$_1^+$ state for the even-$A$ $Ar$ isotopes (from $N=20$ to 30) reasonably well 
and suggest maximum collectivity for the ground state of $^{44}$Ar. However, the 
trend of the excited 0$_2^+$ in even-even Ar isotopes is not that consistent with the 
calculations. The accuracy of the present shell model calculations for predicting the 
0$_2^+$ states will get validation with the experimental observation of the yet unknown 
0$_2^+$ state for $^{48}$Ar in future experimental endeavors.

\section{Acknowledgement}

We thank the NSCL operation team and the A1900 team, especially Tom Ginter, for 
the production and optimization of the secondary beam.  
This work was supported by the U.S. National Science Foundation under 
Grant Nos. PHY-2012522 (FSU), PHY-1848177 (CAREER);   
U.S. Department of Energy, Office of Science, Office of Nuclear Physics
under award Nos. DE-SC0020451 (FRIB), DE-FG02-94ER40848 (UML), DE-AC52-07NA27344 (LLNL),
DE-AC02-06CH11357(ANL) and also by the U.S. Department of Energy
(DOE) National Nuclear Security Administration Grant No. DOE-DE-NA0003906, 
and the Nuclear Science and Security Consortium under Award No. DE-NA0003180.


\begin{thebibliography}{47}%
\makeatletter
\providecommand \@ifxundefined [1]{%
 \@ifx{#1\undefined}
}%
\providecommand \@ifnum [1]{%
 \ifnum #1\expandafter \@firstoftwo
 \else \expandafter \@secondoftwo
 \fi
}%
\providecommand \@ifx [1]{%
 \ifx #1\expandafter \@firstoftwo
 \else \expandafter \@secondoftwo
 \fi
}%
\providecommand \natexlab [1]{#1}%
\providecommand \enquote  [1]{``#1''}%
\providecommand \bibnamefont  [1]{#1}%
\providecommand \bibfnamefont [1]{#1}%
\providecommand \citenamefont [1]{#1}%
\providecommand \href@noop [0]{\@secondoftwo}%
\providecommand \href [0]{\begingroup \@sanitize@url \@href}%
\providecommand \@href[1]{\@@startlink{#1}\@@href}%
\providecommand \@@href[1]{\endgroup#1\@@endlink}%
\providecommand \@sanitize@url [0]{\catcode `\\12\catcode `\$12\catcode
  `\&12\catcode `\#12\catcode `\^12\catcode `\_12\catcode `\%12\relax}%
\providecommand \@@startlink[1]{}%
\providecommand \@@endlink[0]{}%
\providecommand \url  [0]{\begingroup\@sanitize@url \@url }%
\providecommand \@url [1]{\endgroup\@href {#1}{\urlprefix }}%
\providecommand \urlprefix  [0]{URL }%
\providecommand \Eprint [0]{\href }%
\providecommand \doibase [0]{https://doi.org/}%
\providecommand \selectlanguage [0]{\@gobble}%
\providecommand \bibinfo  [0]{\@secondoftwo}%
\providecommand \bibfield  [0]{\@secondoftwo}%
\providecommand \translation [1]{[#1]}%
\providecommand \BibitemOpen [0]{}%
\providecommand \bibitemStop [0]{}%
\providecommand \bibitemNoStop [0]{.\EOS\space}%
\providecommand \EOS [0]{\spacefactor3000\relax}%
\providecommand \BibitemShut  [1]{\csname bibitem#1\endcsname}%
\let\auto@bib@innerbib\@empty
\bibitem [{\citenamefont {Warner}(2004)}]{Warner_magic_number}%
  \BibitemOpen
  \bibfield  {author} {\bibinfo {author} {\bibfnamefont {D.}~\bibnamefont
  {Warner}},\ }\bibfield  {title} {\bibinfo {title} {Not-so-magic numbers},\
  }\href
  {https://EconPapers.repec.org/RePEc:nat:nature:v:430:y:2004:i:6999:d:10.1038_430517a}
  {\bibfield  {journal} {\bibinfo  {journal} {Nature}\ }\textbf {\bibinfo
  {volume} {430}},\ \bibinfo {pages} {517} (\bibinfo {year}
  {2004})}\BibitemShut {NoStop}%
\bibitem [{\citenamefont {Brown}(2001)}]{BROWN2001}%
  \BibitemOpen
  \bibfield  {author} {\bibinfo {author} {\bibfnamefont {B.}~\bibnamefont
  {Brown}},\ }\bibfield  {title} {\bibinfo {title} {The nuclear shell model
  towards the drip lines},\ }\href
  {https://doi.org/https://doi.org/10.1016/S0146-6410(01)00159-4} {\bibfield
  {journal} {\bibinfo  {journal} {Progress in Particle and Nuclear Physics}\
  }\textbf {\bibinfo {volume} {47}},\ \bibinfo {pages} {517} (\bibinfo {year}
  {2001})}\BibitemShut {NoStop}%
\bibitem [{\citenamefont {Sorlin}\ and\ \citenamefont
  {Porquet}(2008)}]{Sorlin_newmagicnumbers}%
  \BibitemOpen
  \bibfield  {author} {\bibinfo {author} {\bibfnamefont {O.}~\bibnamefont
  {Sorlin}}\ and\ \bibinfo {author} {\bibfnamefont {M.-G.}\ \bibnamefont
  {Porquet}},\ }\bibfield  {title} {\bibinfo {title} {Nuclear magic numbers:
  New features far from stability},\ }\href
  {https://doi.org/https://doi.org/10.1016/j.ppnp.2008.05.001} {\bibfield
  {journal} {\bibinfo  {journal} {Progress in Particle and Nuclear Physics}\
  }\textbf {\bibinfo {volume} {61}},\ \bibinfo {pages} {602} (\bibinfo {year}
  {2008})}\BibitemShut {NoStop}%
\bibitem [{\citenamefont {Calinescu}\ \emph {et~al.}(2016)\citenamefont
  {Calinescu}, \citenamefont {C\'aceres}, \citenamefont {Gr\'evy},
  \citenamefont {Sorlin}, \citenamefont {Dombr\'adi}, \citenamefont {Stanoiu},
  \citenamefont {Astabatyan}, \citenamefont {Borcea}, \citenamefont {Borcea},
  \citenamefont {Bowry}, \citenamefont {Catford}, \citenamefont {Cl\'ement},
  \citenamefont {Franchoo}, \citenamefont {Garcia}, \citenamefont {Gillibert},
  \citenamefont {Guerin}, \citenamefont {Kuti}, \citenamefont {Lukyanov},
  \citenamefont {Lepailleur}, \citenamefont {Maslov}, \citenamefont
  {Morfouace}, \citenamefont {Mrazek}, \citenamefont {Negoita}, \citenamefont
  {Niikura}, \citenamefont {Perrot}, \citenamefont {Podoly\'ak}, \citenamefont
  {Petrone}, \citenamefont {Penionzhkevich}, \citenamefont {Roger},
  \citenamefont {Rotaru}, \citenamefont {Sohler}, \citenamefont {Stefan},
  \citenamefont {Thomas}, \citenamefont {Vajta},\ and\ \citenamefont
  {Wilson}}]{46Ar_Calinescu}%
  \BibitemOpen
  \bibfield  {author} {\bibinfo {author} {\bibfnamefont {S.}~\bibnamefont
  {Calinescu}}, \bibinfo {author} {\bibfnamefont {L.}~\bibnamefont
  {C\'aceres}}, \bibinfo {author} {\bibfnamefont {S.}~\bibnamefont {Gr\'evy}},
  \bibinfo {author} {\bibfnamefont {O.}~\bibnamefont {Sorlin}}, \bibinfo
  {author} {\bibfnamefont {Z.}~\bibnamefont {Dombr\'adi}}, \bibinfo {author}
  {\bibfnamefont {M.}~\bibnamefont {Stanoiu}}, \bibinfo {author} {\bibfnamefont
  {R.}~\bibnamefont {Astabatyan}}, \bibinfo {author} {\bibfnamefont
  {C.}~\bibnamefont {Borcea}}, \bibinfo {author} {\bibfnamefont
  {R.}~\bibnamefont {Borcea}}, \bibinfo {author} {\bibfnamefont
  {M.}~\bibnamefont {Bowry}}, \bibinfo {author} {\bibfnamefont
  {W.}~\bibnamefont {Catford}}, \bibinfo {author} {\bibfnamefont
  {E.}~\bibnamefont {Cl\'ement}}, \bibinfo {author} {\bibfnamefont
  {S.}~\bibnamefont {Franchoo}}, \bibinfo {author} {\bibfnamefont
  {R.}~\bibnamefont {Garcia}}, \bibinfo {author} {\bibfnamefont
  {R.}~\bibnamefont {Gillibert}}, \bibinfo {author} {\bibfnamefont {I.~H.}\
  \bibnamefont {Guerin}}, \bibinfo {author} {\bibfnamefont {I.}~\bibnamefont
  {Kuti}}, \bibinfo {author} {\bibfnamefont {S.}~\bibnamefont {Lukyanov}},
  \bibinfo {author} {\bibfnamefont {A.}~\bibnamefont {Lepailleur}}, \bibinfo
  {author} {\bibfnamefont {V.}~\bibnamefont {Maslov}}, \bibinfo {author}
  {\bibfnamefont {P.}~\bibnamefont {Morfouace}}, \bibinfo {author}
  {\bibfnamefont {J.}~\bibnamefont {Mrazek}}, \bibinfo {author} {\bibfnamefont
  {F.}~\bibnamefont {Negoita}}, \bibinfo {author} {\bibfnamefont
  {M.}~\bibnamefont {Niikura}}, \bibinfo {author} {\bibfnamefont
  {L.}~\bibnamefont {Perrot}}, \bibinfo {author} {\bibfnamefont
  {Z.}~\bibnamefont {Podoly\'ak}}, \bibinfo {author} {\bibfnamefont
  {C.}~\bibnamefont {Petrone}}, \bibinfo {author} {\bibfnamefont
  {Y.}~\bibnamefont {Penionzhkevich}}, \bibinfo {author} {\bibfnamefont
  {T.}~\bibnamefont {Roger}}, \bibinfo {author} {\bibfnamefont
  {F.}~\bibnamefont {Rotaru}}, \bibinfo {author} {\bibfnamefont
  {D.}~\bibnamefont {Sohler}}, \bibinfo {author} {\bibfnamefont
  {I.}~\bibnamefont {Stefan}}, \bibinfo {author} {\bibfnamefont {J.~C.}\
  \bibnamefont {Thomas}}, \bibinfo {author} {\bibfnamefont {Z.}~\bibnamefont
  {Vajta}},\ and\ \bibinfo {author} {\bibfnamefont {E.}~\bibnamefont
  {Wilson}},\ }\bibfield  {title} {\bibinfo {title} {Coulomb excitation of
  $^{44}\mathrm{Ca}$ and $^{46}\mathrm{Ar}$},\ }\href
  {https://doi.org/10.1103/PhysRevC.93.044333} {\bibfield  {journal} {\bibinfo
  {journal} {Phys. Rev. C}\ }\textbf {\bibinfo {volume} {93}},\ \bibinfo
  {pages} {044333} (\bibinfo {year} {2016})}\BibitemShut {NoStop}%
\bibitem [{\citenamefont {Bastin}\ \emph {et~al.}(2007)\citenamefont {Bastin},
  \citenamefont {Gr\'evy}, \citenamefont {Sohler}, \citenamefont {Sorlin},
  \citenamefont {Dombr\'adi}, \citenamefont {Achouri}, \citenamefont
  {Ang\'elique}, \citenamefont {Azaiez}, \citenamefont {Baiborodin},
  \citenamefont {Borcea}, \citenamefont {Bourgeois}, \citenamefont {Buta},
  \citenamefont {B\"urger}, \citenamefont {Chapman}, \citenamefont {Dalouzy},
  \citenamefont {Dlouhy}, \citenamefont {Drouard}, \citenamefont {Elekes},
  \citenamefont {Franchoo}, \citenamefont {Iacob}, \citenamefont {Laurent},
  \citenamefont {Lazar}, \citenamefont {Liang}, \citenamefont {Li\'enard},
  \citenamefont {Mrazek}, \citenamefont {Nalpas}, \citenamefont {Negoita},
  \citenamefont {Orr}, \citenamefont {Penionzhkevich}, \citenamefont
  {Podoly\'ak}, \citenamefont {Pougheon}, \citenamefont {Roussel-Chomaz},
  \citenamefont {Saint-Laurent}, \citenamefont {Stanoiu}, \citenamefont
  {Stefan}, \citenamefont {Nowacki},\ and\ \citenamefont
  {Poves}}]{42Si_Bastin}%
  \BibitemOpen
  \bibfield  {author} {\bibinfo {author} {\bibfnamefont {B.}~\bibnamefont
  {Bastin}}, \bibinfo {author} {\bibfnamefont {S.}~\bibnamefont {Gr\'evy}},
  \bibinfo {author} {\bibfnamefont {D.}~\bibnamefont {Sohler}}, \bibinfo
  {author} {\bibfnamefont {O.}~\bibnamefont {Sorlin}}, \bibinfo {author}
  {\bibfnamefont {Z.}~\bibnamefont {Dombr\'adi}}, \bibinfo {author}
  {\bibfnamefont {N.~L.}\ \bibnamefont {Achouri}}, \bibinfo {author}
  {\bibfnamefont {J.~C.}\ \bibnamefont {Ang\'elique}}, \bibinfo {author}
  {\bibfnamefont {F.}~\bibnamefont {Azaiez}}, \bibinfo {author} {\bibfnamefont
  {D.}~\bibnamefont {Baiborodin}}, \bibinfo {author} {\bibfnamefont
  {R.}~\bibnamefont {Borcea}}, \bibinfo {author} {\bibfnamefont
  {C.}~\bibnamefont {Bourgeois}}, \bibinfo {author} {\bibfnamefont
  {A.}~\bibnamefont {Buta}}, \bibinfo {author} {\bibfnamefont {A.}~\bibnamefont
  {B\"urger}}, \bibinfo {author} {\bibfnamefont {R.}~\bibnamefont {Chapman}},
  \bibinfo {author} {\bibfnamefont {J.~C.}\ \bibnamefont {Dalouzy}}, \bibinfo
  {author} {\bibfnamefont {Z.}~\bibnamefont {Dlouhy}}, \bibinfo {author}
  {\bibfnamefont {A.}~\bibnamefont {Drouard}}, \bibinfo {author} {\bibfnamefont
  {Z.}~\bibnamefont {Elekes}}, \bibinfo {author} {\bibfnamefont
  {S.}~\bibnamefont {Franchoo}}, \bibinfo {author} {\bibfnamefont
  {S.}~\bibnamefont {Iacob}}, \bibinfo {author} {\bibfnamefont
  {B.}~\bibnamefont {Laurent}}, \bibinfo {author} {\bibfnamefont
  {M.}~\bibnamefont {Lazar}}, \bibinfo {author} {\bibfnamefont
  {X.}~\bibnamefont {Liang}}, \bibinfo {author} {\bibfnamefont
  {E.}~\bibnamefont {Li\'enard}}, \bibinfo {author} {\bibfnamefont
  {J.}~\bibnamefont {Mrazek}}, \bibinfo {author} {\bibfnamefont
  {L.}~\bibnamefont {Nalpas}}, \bibinfo {author} {\bibfnamefont
  {F.}~\bibnamefont {Negoita}}, \bibinfo {author} {\bibfnamefont {N.~A.}\
  \bibnamefont {Orr}}, \bibinfo {author} {\bibfnamefont {Y.}~\bibnamefont
  {Penionzhkevich}}, \bibinfo {author} {\bibfnamefont {Z.}~\bibnamefont
  {Podoly\'ak}}, \bibinfo {author} {\bibfnamefont {F.}~\bibnamefont
  {Pougheon}}, \bibinfo {author} {\bibfnamefont {P.}~\bibnamefont
  {Roussel-Chomaz}}, \bibinfo {author} {\bibfnamefont {M.~G.}\ \bibnamefont
  {Saint-Laurent}}, \bibinfo {author} {\bibfnamefont {M.}~\bibnamefont
  {Stanoiu}}, \bibinfo {author} {\bibfnamefont {I.}~\bibnamefont {Stefan}},
  \bibinfo {author} {\bibfnamefont {F.}~\bibnamefont {Nowacki}},\ and\ \bibinfo
  {author} {\bibfnamefont {A.}~\bibnamefont {Poves}},\ }\bibfield  {title}
  {\bibinfo {title} {Collapse of the $n=28$ shell closure in
  $^{42}\mathrm{S}\mathrm{i}$},\ }\href
  {https://doi.org/10.1103/PhysRevLett.99.022503} {\bibfield  {journal}
  {\bibinfo  {journal} {Phys. Rev. Lett.}\ }\textbf {\bibinfo {volume} {99}},\
  \bibinfo {pages} {022503} (\bibinfo {year} {2007})}\BibitemShut {NoStop}%
\bibitem [{\citenamefont {Force}\ \emph {et~al.}(2010)\citenamefont {Force},
  \citenamefont {Gr\'evy}, \citenamefont {Gaudefroy}, \citenamefont {Sorlin},
  \citenamefont {C\'aceres}, \citenamefont {Rotaru}, \citenamefont {Mrazek},
  \citenamefont {Achouri}, \citenamefont {Ang\'elique}, \citenamefont {Azaiez},
  \citenamefont {Bastin}, \citenamefont {Borcea}, \citenamefont {Buta},
  \citenamefont {Daugas}, \citenamefont {Dlouhy}, \citenamefont {Dombr\'adi},
  \citenamefont {De~Oliveira}, \citenamefont {Negoita}, \citenamefont
  {Penionzhkevich}, \citenamefont {Saint-Laurent}, \citenamefont {Sohler},
  \citenamefont {Stanoiu}, \citenamefont {Stefan}, \citenamefont {Stodel},\
  and\ \citenamefont {Nowacki}}]{44S_Force2010}%
  \BibitemOpen
  \bibfield  {author} {\bibinfo {author} {\bibfnamefont {C.}~\bibnamefont
  {Force}}, \bibinfo {author} {\bibfnamefont {S.}~\bibnamefont {Gr\'evy}},
  \bibinfo {author} {\bibfnamefont {L.}~\bibnamefont {Gaudefroy}}, \bibinfo
  {author} {\bibfnamefont {O.}~\bibnamefont {Sorlin}}, \bibinfo {author}
  {\bibfnamefont {L.}~\bibnamefont {C\'aceres}}, \bibinfo {author}
  {\bibfnamefont {F.}~\bibnamefont {Rotaru}}, \bibinfo {author} {\bibfnamefont
  {J.}~\bibnamefont {Mrazek}}, \bibinfo {author} {\bibfnamefont {N.~L.}\
  \bibnamefont {Achouri}}, \bibinfo {author} {\bibfnamefont {J.~C.}\
  \bibnamefont {Ang\'elique}}, \bibinfo {author} {\bibfnamefont
  {F.}~\bibnamefont {Azaiez}}, \bibinfo {author} {\bibfnamefont
  {B.}~\bibnamefont {Bastin}}, \bibinfo {author} {\bibfnamefont
  {R.}~\bibnamefont {Borcea}}, \bibinfo {author} {\bibfnamefont
  {A.}~\bibnamefont {Buta}}, \bibinfo {author} {\bibfnamefont {J.~M.}\
  \bibnamefont {Daugas}}, \bibinfo {author} {\bibfnamefont {Z.}~\bibnamefont
  {Dlouhy}}, \bibinfo {author} {\bibfnamefont {Z.}~\bibnamefont {Dombr\'adi}},
  \bibinfo {author} {\bibfnamefont {F.}~\bibnamefont {De~Oliveira}}, \bibinfo
  {author} {\bibfnamefont {F.}~\bibnamefont {Negoita}}, \bibinfo {author}
  {\bibfnamefont {Y.}~\bibnamefont {Penionzhkevich}}, \bibinfo {author}
  {\bibfnamefont {M.~G.}\ \bibnamefont {Saint-Laurent}}, \bibinfo {author}
  {\bibfnamefont {D.}~\bibnamefont {Sohler}}, \bibinfo {author} {\bibfnamefont
  {M.}~\bibnamefont {Stanoiu}}, \bibinfo {author} {\bibfnamefont
  {I.}~\bibnamefont {Stefan}}, \bibinfo {author} {\bibfnamefont
  {C.}~\bibnamefont {Stodel}},\ and\ \bibinfo {author} {\bibfnamefont
  {F.}~\bibnamefont {Nowacki}},\ }\bibfield  {title} {\bibinfo {title}
  {Prolate-spherical shape coexistence at $n=28$ in $^{44}\mathbf{S}$},\ }\href
  {https://doi.org/10.1103/PhysRevLett.105.102501} {\bibfield  {journal}
  {\bibinfo  {journal} {Phys. Rev. Lett.}\ }\textbf {\bibinfo {volume} {105}},\
  \bibinfo {pages} {102501} (\bibinfo {year} {2010})}\BibitemShut {NoStop}%
\bibitem [{\citenamefont {Parker}\ \emph {et~al.}(2017)\citenamefont {Parker},
  \citenamefont {Wiedenh\"over}, \citenamefont {Cottle}, \citenamefont {Baker},
  \citenamefont {McPherson}, \citenamefont {Riley}, \citenamefont
  {Santiago-Gonzalez}, \citenamefont {Volya}, \citenamefont {Bader},
  \citenamefont {Baugher}, \citenamefont {Bazin}, \citenamefont {Gade},
  \citenamefont {Ginter}, \citenamefont {Iwasaki}, \citenamefont {Loelius},
  \citenamefont {Morse}, \citenamefont {Recchia}, \citenamefont {Smalley},
  \citenamefont {Stroberg}, \citenamefont {Whitmore}, \citenamefont
  {Weisshaar}, \citenamefont {Lemasson}, \citenamefont {Crawford},
  \citenamefont {Macchiavelli},\ and\ \citenamefont
  {Wimmer}}]{44S_isomer_Parker}%
  \BibitemOpen
  \bibfield  {author} {\bibinfo {author} {\bibfnamefont {J.~J.}\ \bibnamefont
  {Parker}}, \bibinfo {author} {\bibfnamefont {I.}~\bibnamefont
  {Wiedenh\"over}}, \bibinfo {author} {\bibfnamefont {P.~D.}\ \bibnamefont
  {Cottle}}, \bibinfo {author} {\bibfnamefont {J.}~\bibnamefont {Baker}},
  \bibinfo {author} {\bibfnamefont {D.}~\bibnamefont {McPherson}}, \bibinfo
  {author} {\bibfnamefont {M.~A.}\ \bibnamefont {Riley}}, \bibinfo {author}
  {\bibfnamefont {D.}~\bibnamefont {Santiago-Gonzalez}}, \bibinfo {author}
  {\bibfnamefont {A.}~\bibnamefont {Volya}}, \bibinfo {author} {\bibfnamefont
  {V.~M.}\ \bibnamefont {Bader}}, \bibinfo {author} {\bibfnamefont
  {T.}~\bibnamefont {Baugher}}, \bibinfo {author} {\bibfnamefont
  {D.}~\bibnamefont {Bazin}}, \bibinfo {author} {\bibfnamefont
  {A.}~\bibnamefont {Gade}}, \bibinfo {author} {\bibfnamefont {T.}~\bibnamefont
  {Ginter}}, \bibinfo {author} {\bibfnamefont {H.}~\bibnamefont {Iwasaki}},
  \bibinfo {author} {\bibfnamefont {C.}~\bibnamefont {Loelius}}, \bibinfo
  {author} {\bibfnamefont {C.}~\bibnamefont {Morse}}, \bibinfo {author}
  {\bibfnamefont {F.}~\bibnamefont {Recchia}}, \bibinfo {author} {\bibfnamefont
  {D.}~\bibnamefont {Smalley}}, \bibinfo {author} {\bibfnamefont {S.~R.}\
  \bibnamefont {Stroberg}}, \bibinfo {author} {\bibfnamefont {K.}~\bibnamefont
  {Whitmore}}, \bibinfo {author} {\bibfnamefont {D.}~\bibnamefont {Weisshaar}},
  \bibinfo {author} {\bibfnamefont {A.}~\bibnamefont {Lemasson}}, \bibinfo
  {author} {\bibfnamefont {H.~L.}\ \bibnamefont {Crawford}}, \bibinfo {author}
  {\bibfnamefont {A.~O.}\ \bibnamefont {Macchiavelli}},\ and\ \bibinfo {author}
  {\bibfnamefont {K.}~\bibnamefont {Wimmer}},\ }\bibfield  {title} {\bibinfo
  {title} {Isomeric character of the lowest observed ${4}^{+}$ state in
  $^{44}\mathrm{S}$},\ }\href {https://doi.org/10.1103/PhysRevLett.118.052501}
  {\bibfield  {journal} {\bibinfo  {journal} {Phys. Rev. Lett.}\ }\textbf
  {\bibinfo {volume} {118}},\ \bibinfo {pages} {052501} (\bibinfo {year}
  {2017})}\BibitemShut {NoStop}%
\bibitem [{\citenamefont {Otsuka}\ \emph {et~al.}(2020)\citenamefont {Otsuka},
  \citenamefont {Gade}, \citenamefont {Sorlin}, \citenamefont {Suzuki},\ and\
  \citenamefont {Utsuno}}]{shell_evolution}%
  \BibitemOpen
  \bibfield  {author} {\bibinfo {author} {\bibfnamefont {T.}~\bibnamefont
  {Otsuka}}, \bibinfo {author} {\bibfnamefont {A.}~\bibnamefont {Gade}},
  \bibinfo {author} {\bibfnamefont {O.}~\bibnamefont {Sorlin}}, \bibinfo
  {author} {\bibfnamefont {T.}~\bibnamefont {Suzuki}},\ and\ \bibinfo {author}
  {\bibfnamefont {Y.}~\bibnamefont {Utsuno}},\ }\bibfield  {title} {\bibinfo
  {title} {Evolution of shell structure in exotic nuclei},\ }\href
  {https://doi.org/10.1103/RevModPhys.92.015002} {\bibfield  {journal}
  {\bibinfo  {journal} {Rev. Mod. Phys.}\ }\textbf {\bibinfo {volume} {92}},\
  \bibinfo {pages} {015002} (\bibinfo {year} {2020})}\BibitemShut {NoStop}%
\bibitem [{\citenamefont {Glasmacher}\ \emph {et~al.}(1997)\citenamefont
  {Glasmacher}, \citenamefont {Brown}, \citenamefont {Chromik}, \citenamefont
  {Cottle}, \citenamefont {Fauerbach}, \citenamefont {Ibbotson}, \citenamefont
  {Kemper}, \citenamefont {Morrissey}, \citenamefont {Scheit}, \citenamefont
  {Sklenicka},\ and\ \citenamefont {Steiner}}]{Glasmacher_44S}%
  \BibitemOpen
  \bibfield  {author} {\bibinfo {author} {\bibfnamefont {T.}~\bibnamefont
  {Glasmacher}}, \bibinfo {author} {\bibfnamefont {B.}~\bibnamefont {Brown}},
  \bibinfo {author} {\bibfnamefont {M.}~\bibnamefont {Chromik}}, \bibinfo
  {author} {\bibfnamefont {P.}~\bibnamefont {Cottle}}, \bibinfo {author}
  {\bibfnamefont {M.}~\bibnamefont {Fauerbach}}, \bibinfo {author}
  {\bibfnamefont {R.}~\bibnamefont {Ibbotson}}, \bibinfo {author}
  {\bibfnamefont {K.}~\bibnamefont {Kemper}}, \bibinfo {author} {\bibfnamefont
  {D.}~\bibnamefont {Morrissey}}, \bibinfo {author} {\bibfnamefont
  {H.}~\bibnamefont {Scheit}}, \bibinfo {author} {\bibfnamefont
  {D.}~\bibnamefont {Sklenicka}},\ and\ \bibinfo {author} {\bibfnamefont
  {M.}~\bibnamefont {Steiner}},\ }\bibfield  {title} {\bibinfo {title}
  {Collectivity in 44s},\ }\href
  {https://doi.org/https://doi.org/10.1016/S0370-2693(97)00077-4} {\bibfield
  {journal} {\bibinfo  {journal} {Physics Letters B}\ }\textbf {\bibinfo
  {volume} {395}},\ \bibinfo {pages} {163} (\bibinfo {year}
  {1997})}\BibitemShut {NoStop}%
\bibitem [{\citenamefont {Werner}\ \emph {et~al.}(1994)\citenamefont {Werner},
  \citenamefont {Sheikh}, \citenamefont {Nazarewicz}, \citenamefont {Strayer},
  \citenamefont {Umar},\ and\ \citenamefont {Misu}}]{44S_shape_warner}%
  \BibitemOpen
  \bibfield  {author} {\bibinfo {author} {\bibfnamefont {T.}~\bibnamefont
  {Werner}}, \bibinfo {author} {\bibfnamefont {J.}~\bibnamefont {Sheikh}},
  \bibinfo {author} {\bibfnamefont {W.}~\bibnamefont {Nazarewicz}}, \bibinfo
  {author} {\bibfnamefont {M.}~\bibnamefont {Strayer}}, \bibinfo {author}
  {\bibfnamefont {A.}~\bibnamefont {Umar}},\ and\ \bibinfo {author}
  {\bibfnamefont {M.}~\bibnamefont {Misu}},\ }\bibfield  {title} {\bibinfo
  {title} {Shape coexistence around 1644s28: the deformed n = 28 region 1565},\
  }\href {https://doi.org/https://doi.org/10.1016/0370-2693(94)90347-6}
  {\bibfield  {journal} {\bibinfo  {journal} {Physics Letters B}\ }\textbf
  {\bibinfo {volume} {335}},\ \bibinfo {pages} {259} (\bibinfo {year}
  {1994})}\BibitemShut {NoStop}%
\bibitem [{\citenamefont {Longfellow}\ \emph {et~al.}(2021)\citenamefont
  {Longfellow}, \citenamefont {Weisshaar}, \citenamefont {Gade}, \citenamefont
  {Brown}, \citenamefont {Bazin}, \citenamefont {Brown}, \citenamefont {Elman},
  \citenamefont {Pereira}, \citenamefont {Rhodes},\ and\ \citenamefont
  {Spieker}}]{longfellow_44S_21}%
  \BibitemOpen
  \bibfield  {author} {\bibinfo {author} {\bibfnamefont {B.}~\bibnamefont
  {Longfellow}}, \bibinfo {author} {\bibfnamefont {D.}~\bibnamefont
  {Weisshaar}}, \bibinfo {author} {\bibfnamefont {A.}~\bibnamefont {Gade}},
  \bibinfo {author} {\bibfnamefont {B.~A.}\ \bibnamefont {Brown}}, \bibinfo
  {author} {\bibfnamefont {D.}~\bibnamefont {Bazin}}, \bibinfo {author}
  {\bibfnamefont {K.~W.}\ \bibnamefont {Brown}}, \bibinfo {author}
  {\bibfnamefont {B.}~\bibnamefont {Elman}}, \bibinfo {author} {\bibfnamefont
  {J.}~\bibnamefont {Pereira}}, \bibinfo {author} {\bibfnamefont
  {D.}~\bibnamefont {Rhodes}},\ and\ \bibinfo {author} {\bibfnamefont
  {M.}~\bibnamefont {Spieker}},\ }\bibfield  {title} {\bibinfo {title}
  {Quadrupole collectivity in the neutron-rich sulfur isotopes
  $^{38,40,42,44}\mathrm{S}$},\ }\href
  {https://doi.org/10.1103/PhysRevC.103.054309} {\bibfield  {journal} {\bibinfo
   {journal} {Phys. Rev. C}\ }\textbf {\bibinfo {volume} {103}},\ \bibinfo
  {pages} {054309} (\bibinfo {year} {2021})}\BibitemShut {NoStop}%
\bibitem [{\citenamefont {Gade}\ \emph {et~al.}(2006)\citenamefont {Gade},
  \citenamefont {Brown}, \citenamefont {Bazin}, \citenamefont {Campbell},
  \citenamefont {Church}, \citenamefont {Dinca}, \citenamefont {Enders},
  \citenamefont {Glasmacher}, \citenamefont {Horoi}, \citenamefont {Hu},
  \citenamefont {Kemper}, \citenamefont {Mueller}, \citenamefont {Otsuka},
  \citenamefont {Riley}, \citenamefont {Roeder}, \citenamefont {Suzuki},
  \citenamefont {Terry}, \citenamefont {Yurkewicz},\ and\ \citenamefont
  {Zwahlen}}]{Gade_K_Cl_P_ProtonOrbital}%
  \BibitemOpen
  \bibfield  {author} {\bibinfo {author} {\bibfnamefont {A.}~\bibnamefont
  {Gade}}, \bibinfo {author} {\bibfnamefont {B.~A.}\ \bibnamefont {Brown}},
  \bibinfo {author} {\bibfnamefont {D.}~\bibnamefont {Bazin}}, \bibinfo
  {author} {\bibfnamefont {C.~M.}\ \bibnamefont {Campbell}}, \bibinfo {author}
  {\bibfnamefont {J.~A.}\ \bibnamefont {Church}}, \bibinfo {author}
  {\bibfnamefont {D.~C.}\ \bibnamefont {Dinca}}, \bibinfo {author}
  {\bibfnamefont {J.}~\bibnamefont {Enders}}, \bibinfo {author} {\bibfnamefont
  {T.}~\bibnamefont {Glasmacher}}, \bibinfo {author} {\bibfnamefont
  {M.}~\bibnamefont {Horoi}}, \bibinfo {author} {\bibfnamefont
  {Z.}~\bibnamefont {Hu}}, \bibinfo {author} {\bibfnamefont {K.~W.}\
  \bibnamefont {Kemper}}, \bibinfo {author} {\bibfnamefont {W.~F.}\
  \bibnamefont {Mueller}}, \bibinfo {author} {\bibfnamefont {T.}~\bibnamefont
  {Otsuka}}, \bibinfo {author} {\bibfnamefont {L.~A.}\ \bibnamefont {Riley}},
  \bibinfo {author} {\bibfnamefont {B.~T.}\ \bibnamefont {Roeder}}, \bibinfo
  {author} {\bibfnamefont {T.}~\bibnamefont {Suzuki}}, \bibinfo {author}
  {\bibfnamefont {J.~R.}\ \bibnamefont {Terry}}, \bibinfo {author}
  {\bibfnamefont {K.~L.}\ \bibnamefont {Yurkewicz}},\ and\ \bibinfo {author}
  {\bibfnamefont {H.}~\bibnamefont {Zwahlen}},\ }\bibfield  {title} {\bibinfo
  {title} {Evolution of the $e(1/{2}_{1}^{+})\ensuremath{-}e(3/{2}_{1}^{+})$
  energy spacing in odd-mass k, cl, and p isotopes for
  $n=20\text{\ensuremath{-}}28$},\ }\href
  {https://doi.org/10.1103/PhysRevC.74.034322} {\bibfield  {journal} {\bibinfo
  {journal} {Phys. Rev. C}\ }\textbf {\bibinfo {volume} {74}},\ \bibinfo
  {pages} {034322} (\bibinfo {year} {2006})}\BibitemShut {NoStop}%
\bibitem [{\citenamefont {Otsuka}\ \emph {et~al.}(2005)\citenamefont {Otsuka},
  \citenamefont {Suzuki}, \citenamefont {Fujimoto}, \citenamefont {Grawe},\
  and\ \citenamefont {Akaishi}}]{Otsuka_tensor_2005}%
  \BibitemOpen
  \bibfield  {author} {\bibinfo {author} {\bibfnamefont {T.}~\bibnamefont
  {Otsuka}}, \bibinfo {author} {\bibfnamefont {T.}~\bibnamefont {Suzuki}},
  \bibinfo {author} {\bibfnamefont {R.}~\bibnamefont {Fujimoto}}, \bibinfo
  {author} {\bibfnamefont {H.}~\bibnamefont {Grawe}},\ and\ \bibinfo {author}
  {\bibfnamefont {Y.}~\bibnamefont {Akaishi}},\ }\bibfield  {title} {\bibinfo
  {title} {Evolution of nuclear shells due to the tensor force},\ }\href
  {https://doi.org/10.1103/PhysRevLett.95.232502} {\bibfield  {journal}
  {\bibinfo  {journal} {Phys. Rev. Lett.}\ }\textbf {\bibinfo {volume} {95}},\
  \bibinfo {pages} {232502} (\bibinfo {year} {2005})}\BibitemShut {NoStop}%
\bibitem [{\citenamefont {Stroberg}\ \emph {et~al.}(2012)\citenamefont
  {Stroberg}, \citenamefont {Gade}, \citenamefont {Baugher}, \citenamefont
  {Bazin}, \citenamefont {Brown}, \citenamefont {Cook}, \citenamefont
  {Glasmacher}, \citenamefont {Grinyer}, \citenamefont {McDaniel},
  \citenamefont {Ratkiewicz},\ and\ \citenamefont {Weisshaar}}]{stroberg}%
  \BibitemOpen
  \bibfield  {author} {\bibinfo {author} {\bibfnamefont {S.~R.}\ \bibnamefont
  {Stroberg}}, \bibinfo {author} {\bibfnamefont {A.}~\bibnamefont {Gade}},
  \bibinfo {author} {\bibfnamefont {T.}~\bibnamefont {Baugher}}, \bibinfo
  {author} {\bibfnamefont {D.}~\bibnamefont {Bazin}}, \bibinfo {author}
  {\bibfnamefont {B.~A.}\ \bibnamefont {Brown}}, \bibinfo {author}
  {\bibfnamefont {J.~M.}\ \bibnamefont {Cook}}, \bibinfo {author}
  {\bibfnamefont {T.}~\bibnamefont {Glasmacher}}, \bibinfo {author}
  {\bibfnamefont {G.~F.}\ \bibnamefont {Grinyer}}, \bibinfo {author}
  {\bibfnamefont {S.}~\bibnamefont {McDaniel}}, \bibinfo {author}
  {\bibfnamefont {A.}~\bibnamefont {Ratkiewicz}},\ and\ \bibinfo {author}
  {\bibfnamefont {D.}~\bibnamefont {Weisshaar}},\ }\bibfield  {title} {\bibinfo
  {title} {In-beam $\ensuremath{\gamma}$-ray spectroscopy of
  ${}^{43\ensuremath{-}46}$cl},\ }\href
  {https://doi.org/10.1103/PhysRevC.86.024321} {\bibfield  {journal} {\bibinfo
  {journal} {Phys. Rev. C}\ }\textbf {\bibinfo {volume} {86}},\ \bibinfo
  {pages} {024321} (\bibinfo {year} {2012})}\BibitemShut {NoStop}%
\bibitem [{\citenamefont {Blaum}\ \emph {et~al.}(2008)\citenamefont {Blaum},
  \citenamefont {Geithner}, \citenamefont {Lassen}, \citenamefont {Lievens},
  \citenamefont {Marinova},\ and\ \citenamefont {Neugart}}]{43Ar_BLAUM2008}%
  \BibitemOpen
  \bibfield  {author} {\bibinfo {author} {\bibfnamefont {K.}~\bibnamefont
  {Blaum}}, \bibinfo {author} {\bibfnamefont {W.}~\bibnamefont {Geithner}},
  \bibinfo {author} {\bibfnamefont {J.}~\bibnamefont {Lassen}}, \bibinfo
  {author} {\bibfnamefont {P.}~\bibnamefont {Lievens}}, \bibinfo {author}
  {\bibfnamefont {K.}~\bibnamefont {Marinova}},\ and\ \bibinfo {author}
  {\bibfnamefont {R.}~\bibnamefont {Neugart}},\ }\bibfield  {title} {\bibinfo
  {title} {Nuclear moments and charge radii of argon isotopes between the
  neutron-shell closures n=20 and n=28},\ }\href
  {https://doi.org/https://doi.org/10.1016/j.nuclphysa.2007.11.004} {\bibfield
  {journal} {\bibinfo  {journal} {Nuclear Physics A}\ }\textbf {\bibinfo
  {volume} {799}},\ \bibinfo {pages} {30} (\bibinfo {year} {2008})}\BibitemShut
  {NoStop}%
\bibitem [{\citenamefont {Lu}\ \emph {et~al.}(2013)\citenamefont {Lu},
  \citenamefont {Lee}, \citenamefont {Tsang}, \citenamefont {Bazin},
  \citenamefont {Coupland}, \citenamefont {Henzl}, \citenamefont {Henzlova},
  \citenamefont {Kilburn}, \citenamefont {Lynch}, \citenamefont {Rogers},
  \citenamefont {Sanetullaev}, \citenamefont {Sun}, \citenamefont {Youngs},
  \citenamefont {Charity}, \citenamefont {Sobotka}, \citenamefont {Famiano},
  \citenamefont {Hudan}, \citenamefont {Horoi},\ and\ \citenamefont
  {Ye}}]{F_Lu_2013}%
  \BibitemOpen
  \bibfield  {author} {\bibinfo {author} {\bibfnamefont {F.}~\bibnamefont
  {Lu}}, \bibinfo {author} {\bibfnamefont {J.}~\bibnamefont {Lee}}, \bibinfo
  {author} {\bibfnamefont {M.~B.}\ \bibnamefont {Tsang}}, \bibinfo {author}
  {\bibfnamefont {D.}~\bibnamefont {Bazin}}, \bibinfo {author} {\bibfnamefont
  {D.}~\bibnamefont {Coupland}}, \bibinfo {author} {\bibfnamefont
  {V.}~\bibnamefont {Henzl}}, \bibinfo {author} {\bibfnamefont
  {D.}~\bibnamefont {Henzlova}}, \bibinfo {author} {\bibfnamefont
  {M.}~\bibnamefont {Kilburn}}, \bibinfo {author} {\bibfnamefont {W.~G.}\
  \bibnamefont {Lynch}}, \bibinfo {author} {\bibfnamefont {A.~M.}\ \bibnamefont
  {Rogers}}, \bibinfo {author} {\bibfnamefont {A.}~\bibnamefont {Sanetullaev}},
  \bibinfo {author} {\bibfnamefont {Z.~Y.}\ \bibnamefont {Sun}}, \bibinfo
  {author} {\bibfnamefont {M.}~\bibnamefont {Youngs}}, \bibinfo {author}
  {\bibfnamefont {R.~J.}\ \bibnamefont {Charity}}, \bibinfo {author}
  {\bibfnamefont {L.~G.}\ \bibnamefont {Sobotka}}, \bibinfo {author}
  {\bibfnamefont {M.}~\bibnamefont {Famiano}}, \bibinfo {author} {\bibfnamefont
  {S.}~\bibnamefont {Hudan}}, \bibinfo {author} {\bibfnamefont
  {M.}~\bibnamefont {Horoi}},\ and\ \bibinfo {author} {\bibfnamefont {Y.~L.}\
  \bibnamefont {Ye}},\ }\bibfield  {title} {\bibinfo {title} {Neutron-hole
  states in ${}^{45}$ar from ${}^{1}$h(${}^{46}$ar, $d$) ${}^{45}$ar
  reactions},\ }\href {https://doi.org/10.1103/PhysRevC.88.017604} {\bibfield
  {journal} {\bibinfo  {journal} {Phys. Rev. C}\ }\textbf {\bibinfo {volume}
  {88}},\ \bibinfo {pages} {017604} (\bibinfo {year} {2013})}\BibitemShut
  {NoStop}%
\bibitem [{\citenamefont {Zieli\ifmmode~\acute{n}\else \'{n}\fi{}ska}\ \emph
  {et~al.}(2009)\citenamefont {Zieli\ifmmode~\acute{n}\else \'{n}\fi{}ska},
  \citenamefont {G\"orgen}, \citenamefont {Cl\'ement}, \citenamefont
  {Delaroche}, \citenamefont {Girod}, \citenamefont {Korten}, \citenamefont
  {B\"urger}, \citenamefont {Catford}, \citenamefont {Dossat}, \citenamefont
  {Iwanicki}, \citenamefont {Libert}, \citenamefont {Ljungvall}, \citenamefont
  {Napiorkowski}, \citenamefont {Obertelli}, \citenamefont {Pietak},
  \citenamefont {Rodr\'{\i}guez-Guzm\'an}, \citenamefont {Sletten},
  \citenamefont {Srebrny}, \citenamefont {Theisen},\ and\ \citenamefont
  {Wrzosek}}]{44Ar_coulomb}%
  \BibitemOpen
  \bibfield  {author} {\bibinfo {author} {\bibfnamefont {M.}~\bibnamefont
  {Zieli\ifmmode~\acute{n}\else \'{n}\fi{}ska}}, \bibinfo {author}
  {\bibfnamefont {A.}~\bibnamefont {G\"orgen}}, \bibinfo {author}
  {\bibfnamefont {E.}~\bibnamefont {Cl\'ement}}, \bibinfo {author}
  {\bibfnamefont {J.~P.}\ \bibnamefont {Delaroche}}, \bibinfo {author}
  {\bibfnamefont {M.}~\bibnamefont {Girod}}, \bibinfo {author} {\bibfnamefont
  {W.}~\bibnamefont {Korten}}, \bibinfo {author} {\bibfnamefont
  {A.}~\bibnamefont {B\"urger}}, \bibinfo {author} {\bibfnamefont
  {W.}~\bibnamefont {Catford}}, \bibinfo {author} {\bibfnamefont
  {C.}~\bibnamefont {Dossat}}, \bibinfo {author} {\bibfnamefont
  {J.}~\bibnamefont {Iwanicki}}, \bibinfo {author} {\bibfnamefont
  {J.}~\bibnamefont {Libert}}, \bibinfo {author} {\bibfnamefont
  {J.}~\bibnamefont {Ljungvall}}, \bibinfo {author} {\bibfnamefont {P.~J.}\
  \bibnamefont {Napiorkowski}}, \bibinfo {author} {\bibfnamefont
  {A.}~\bibnamefont {Obertelli}}, \bibinfo {author} {\bibfnamefont
  {D.}~\bibnamefont {Pietak}}, \bibinfo {author} {\bibfnamefont
  {R.}~\bibnamefont {Rodr\'{\i}guez-Guzm\'an}}, \bibinfo {author}
  {\bibfnamefont {G.}~\bibnamefont {Sletten}}, \bibinfo {author} {\bibfnamefont
  {J.}~\bibnamefont {Srebrny}}, \bibinfo {author} {\bibfnamefont
  {C.}~\bibnamefont {Theisen}},\ and\ \bibinfo {author} {\bibfnamefont
  {K.}~\bibnamefont {Wrzosek}},\ }\bibfield  {title} {\bibinfo {title} {Shape
  of $^{44}\mathrm{Ar}$: Onset of deformation in neutron-rich nuclei near
  $^{48}\mathrm{Ca}$},\ }\href {https://doi.org/10.1103/PhysRevC.80.014317}
  {\bibfield  {journal} {\bibinfo  {journal} {Phys. Rev. C}\ }\textbf {\bibinfo
  {volume} {80}},\ \bibinfo {pages} {014317} (\bibinfo {year}
  {2009})}\BibitemShut {NoStop}%
\bibitem [{\citenamefont {Scheit}\ \emph {et~al.}(1996)\citenamefont {Scheit},
  \citenamefont {Glasmacher}, \citenamefont {Brown}, \citenamefont {Brown},
  \citenamefont {Cottle}, \citenamefont {Hansen}, \citenamefont {Harkewicz},
  \citenamefont {Hellstr\"om}, \citenamefont {Ibbotson}, \citenamefont
  {Jewell}, \citenamefont {Kemper}, \citenamefont {Morrissey}, \citenamefont
  {Steiner}, \citenamefont {Thirolf},\ and\ \citenamefont
  {Thoennessen}}]{S_Ar_deformation_Scheit}%
  \BibitemOpen
  \bibfield  {author} {\bibinfo {author} {\bibfnamefont {H.}~\bibnamefont
  {Scheit}}, \bibinfo {author} {\bibfnamefont {T.}~\bibnamefont {Glasmacher}},
  \bibinfo {author} {\bibfnamefont {B.~A.}\ \bibnamefont {Brown}}, \bibinfo
  {author} {\bibfnamefont {J.~A.}\ \bibnamefont {Brown}}, \bibinfo {author}
  {\bibfnamefont {P.~D.}\ \bibnamefont {Cottle}}, \bibinfo {author}
  {\bibfnamefont {P.~G.}\ \bibnamefont {Hansen}}, \bibinfo {author}
  {\bibfnamefont {R.}~\bibnamefont {Harkewicz}}, \bibinfo {author}
  {\bibfnamefont {M.}~\bibnamefont {Hellstr\"om}}, \bibinfo {author}
  {\bibfnamefont {R.~W.}\ \bibnamefont {Ibbotson}}, \bibinfo {author}
  {\bibfnamefont {J.~K.}\ \bibnamefont {Jewell}}, \bibinfo {author}
  {\bibfnamefont {K.~W.}\ \bibnamefont {Kemper}}, \bibinfo {author}
  {\bibfnamefont {D.~J.}\ \bibnamefont {Morrissey}}, \bibinfo {author}
  {\bibfnamefont {M.}~\bibnamefont {Steiner}}, \bibinfo {author} {\bibfnamefont
  {P.}~\bibnamefont {Thirolf}},\ and\ \bibinfo {author} {\bibfnamefont
  {M.}~\bibnamefont {Thoennessen}},\ }\bibfield  {title} {\bibinfo {title} {New
  region of deformation: The neutron-rich sulfur isotopes},\ }\href
  {https://doi.org/10.1103/PhysRevLett.77.3967} {\bibfield  {journal} {\bibinfo
   {journal} {Phys. Rev. Lett.}\ }\textbf {\bibinfo {volume} {77}},\ \bibinfo
  {pages} {3967} (\bibinfo {year} {1996})}\BibitemShut {NoStop}%
\bibitem [{\citenamefont {Gade}\ \emph {et~al.}(2003)\citenamefont {Gade},
  \citenamefont {Bazin}, \citenamefont {Campbell}, \citenamefont {Church},
  \citenamefont {Dinca}, \citenamefont {Enders}, \citenamefont {Glasmacher},
  \citenamefont {Hu}, \citenamefont {Kemper}, \citenamefont {Mueller},
  \citenamefont {Olliver}, \citenamefont {Perry}, \citenamefont {Riley},
  \citenamefont {Roeder}, \citenamefont {Sherrill},\ and\ \citenamefont
  {Terry}}]{Gade_46Ar_2003}%
  \BibitemOpen
  \bibfield  {author} {\bibinfo {author} {\bibfnamefont {A.}~\bibnamefont
  {Gade}}, \bibinfo {author} {\bibfnamefont {D.}~\bibnamefont {Bazin}},
  \bibinfo {author} {\bibfnamefont {C.~M.}\ \bibnamefont {Campbell}}, \bibinfo
  {author} {\bibfnamefont {J.~A.}\ \bibnamefont {Church}}, \bibinfo {author}
  {\bibfnamefont {D.~C.}\ \bibnamefont {Dinca}}, \bibinfo {author}
  {\bibfnamefont {J.}~\bibnamefont {Enders}}, \bibinfo {author} {\bibfnamefont
  {T.}~\bibnamefont {Glasmacher}}, \bibinfo {author} {\bibfnamefont
  {Z.}~\bibnamefont {Hu}}, \bibinfo {author} {\bibfnamefont {K.~W.}\
  \bibnamefont {Kemper}}, \bibinfo {author} {\bibfnamefont {W.~F.}\
  \bibnamefont {Mueller}}, \bibinfo {author} {\bibfnamefont {H.}~\bibnamefont
  {Olliver}}, \bibinfo {author} {\bibfnamefont {B.~C.}\ \bibnamefont {Perry}},
  \bibinfo {author} {\bibfnamefont {L.~A.}\ \bibnamefont {Riley}}, \bibinfo
  {author} {\bibfnamefont {B.~T.}\ \bibnamefont {Roeder}}, \bibinfo {author}
  {\bibfnamefont {B.~M.}\ \bibnamefont {Sherrill}},\ and\ \bibinfo {author}
  {\bibfnamefont {J.~R.}\ \bibnamefont {Terry}},\ }\bibfield  {title} {\bibinfo
  {title} {Detailed experimental study on intermediate-energy coulomb
  excitation of ${}^{46}\mathrm{Ar}$},\ }\href
  {https://doi.org/10.1103/PhysRevC.68.014302} {\bibfield  {journal} {\bibinfo
  {journal} {Phys. Rev. C}\ }\textbf {\bibinfo {volume} {68}},\ \bibinfo
  {pages} {014302} (\bibinfo {year} {2003})}\BibitemShut {NoStop}%
\bibitem [{\citenamefont {Mengoni}\ \emph {et~al.}(2010)\citenamefont
  {Mengoni}, \citenamefont {Valiente-Dob\'on}, \citenamefont {Gadea},
  \citenamefont {Lunardi}, \citenamefont {Lenzi}, \citenamefont {Broda},
  \citenamefont {Dewald}, \citenamefont {Pissulla}, \citenamefont {Angus},
  \citenamefont {Aydin}, \citenamefont {Bazzacco}, \citenamefont {Benzoni},
  \citenamefont {Bizzeti}, \citenamefont {Bizzeti-Sona}, \citenamefont
  {Boutachkov}, \citenamefont {Corradi}, \citenamefont {Crespi}, \citenamefont
  {de~Angelis}, \citenamefont {Farnea}, \citenamefont {Fioretto}, \citenamefont
  {Goergen}, \citenamefont {Gorska}, \citenamefont {Gottardo}, \citenamefont
  {Grodner}, \citenamefont {Howard}, \citenamefont {Kr\'olas}, \citenamefont
  {Leoni}, \citenamefont {Mason}, \citenamefont {Montanari}, \citenamefont
  {Montagnoli}, \citenamefont {Napoli}, \citenamefont {Obertelli},
  \citenamefont {Orlandi}, \citenamefont {Paw\l{}at}, \citenamefont
  {Pollarolo}, \citenamefont {Recchia}, \citenamefont {Algora}, \citenamefont
  {Rubio}, \citenamefont {Sahin}, \citenamefont {Scarlassara}, \citenamefont
  {Silvestri}, \citenamefont {Smith}, \citenamefont {Stefanini}, \citenamefont
  {Steppenbeck}, \citenamefont {Szilner}, \citenamefont {Ur}, \citenamefont
  {Wady},\ and\ \citenamefont {Wrzesi\ifmmode~\acute{n}\else
  \'{n}\fi{}ski}}]{44-46Ar_Mengoni_2010}%
  \BibitemOpen
  \bibfield  {author} {\bibinfo {author} {\bibfnamefont {D.}~\bibnamefont
  {Mengoni}}, \bibinfo {author} {\bibfnamefont {J.~J.}\ \bibnamefont
  {Valiente-Dob\'on}}, \bibinfo {author} {\bibfnamefont {A.}~\bibnamefont
  {Gadea}}, \bibinfo {author} {\bibfnamefont {S.}~\bibnamefont {Lunardi}},
  \bibinfo {author} {\bibfnamefont {S.~M.}\ \bibnamefont {Lenzi}}, \bibinfo
  {author} {\bibfnamefont {R.}~\bibnamefont {Broda}}, \bibinfo {author}
  {\bibfnamefont {A.}~\bibnamefont {Dewald}}, \bibinfo {author} {\bibfnamefont
  {T.}~\bibnamefont {Pissulla}}, \bibinfo {author} {\bibfnamefont {L.~J.}\
  \bibnamefont {Angus}}, \bibinfo {author} {\bibfnamefont {S.}~\bibnamefont
  {Aydin}}, \bibinfo {author} {\bibfnamefont {D.}~\bibnamefont {Bazzacco}},
  \bibinfo {author} {\bibfnamefont {G.}~\bibnamefont {Benzoni}}, \bibinfo
  {author} {\bibfnamefont {P.~G.}\ \bibnamefont {Bizzeti}}, \bibinfo {author}
  {\bibfnamefont {A.~M.}\ \bibnamefont {Bizzeti-Sona}}, \bibinfo {author}
  {\bibfnamefont {P.}~\bibnamefont {Boutachkov}}, \bibinfo {author}
  {\bibfnamefont {L.}~\bibnamefont {Corradi}}, \bibinfo {author} {\bibfnamefont
  {F.}~\bibnamefont {Crespi}}, \bibinfo {author} {\bibfnamefont
  {G.}~\bibnamefont {de~Angelis}}, \bibinfo {author} {\bibfnamefont
  {E.}~\bibnamefont {Farnea}}, \bibinfo {author} {\bibfnamefont
  {E.}~\bibnamefont {Fioretto}}, \bibinfo {author} {\bibfnamefont
  {A.}~\bibnamefont {Goergen}}, \bibinfo {author} {\bibfnamefont
  {M.}~\bibnamefont {Gorska}}, \bibinfo {author} {\bibfnamefont
  {A.}~\bibnamefont {Gottardo}}, \bibinfo {author} {\bibfnamefont
  {E.}~\bibnamefont {Grodner}}, \bibinfo {author} {\bibfnamefont {A.~M.}\
  \bibnamefont {Howard}}, \bibinfo {author} {\bibfnamefont {W.}~\bibnamefont
  {Kr\'olas}}, \bibinfo {author} {\bibfnamefont {S.}~\bibnamefont {Leoni}},
  \bibinfo {author} {\bibfnamefont {P.}~\bibnamefont {Mason}}, \bibinfo
  {author} {\bibfnamefont {D.}~\bibnamefont {Montanari}}, \bibinfo {author}
  {\bibfnamefont {G.}~\bibnamefont {Montagnoli}}, \bibinfo {author}
  {\bibfnamefont {D.~R.}\ \bibnamefont {Napoli}}, \bibinfo {author}
  {\bibfnamefont {A.}~\bibnamefont {Obertelli}}, \bibinfo {author}
  {\bibfnamefont {R.}~\bibnamefont {Orlandi}}, \bibinfo {author} {\bibfnamefont
  {T.}~\bibnamefont {Paw\l{}at}}, \bibinfo {author} {\bibfnamefont
  {G.}~\bibnamefont {Pollarolo}}, \bibinfo {author} {\bibfnamefont
  {F.}~\bibnamefont {Recchia}}, \bibinfo {author} {\bibfnamefont
  {A.}~\bibnamefont {Algora}}, \bibinfo {author} {\bibfnamefont
  {B.}~\bibnamefont {Rubio}}, \bibinfo {author} {\bibfnamefont
  {E.}~\bibnamefont {Sahin}}, \bibinfo {author} {\bibfnamefont
  {F.}~\bibnamefont {Scarlassara}}, \bibinfo {author} {\bibfnamefont
  {R.}~\bibnamefont {Silvestri}}, \bibinfo {author} {\bibfnamefont {J.~F.}\
  \bibnamefont {Smith}}, \bibinfo {author} {\bibfnamefont {A.~M.}\ \bibnamefont
  {Stefanini}}, \bibinfo {author} {\bibfnamefont {D.}~\bibnamefont
  {Steppenbeck}}, \bibinfo {author} {\bibfnamefont {S.}~\bibnamefont
  {Szilner}}, \bibinfo {author} {\bibfnamefont {C.~A.}\ \bibnamefont {Ur}},
  \bibinfo {author} {\bibfnamefont {P.~T.}\ \bibnamefont {Wady}},\ and\
  \bibinfo {author} {\bibfnamefont {J.}~\bibnamefont
  {Wrzesi\ifmmode~\acute{n}\else \'{n}\fi{}ski}},\ }\bibfield  {title}
  {\bibinfo {title} {Lifetime measurements of excited states in neutron-rich
  $^{44,46}\mathrm{Ar}$ populated via a multinucleon transfer reaction},\
  }\href {https://doi.org/10.1103/PhysRevC.82.024308} {\bibfield  {journal}
  {\bibinfo  {journal} {Phys. Rev. C}\ }\textbf {\bibinfo {volume} {82}},\
  \bibinfo {pages} {024308} (\bibinfo {year} {2010})}\BibitemShut {NoStop}%
\bibitem [{\citenamefont {Gade}\ and\ \citenamefont {Sherill}(2016)}]{brad}%
  \BibitemOpen
  \bibfield  {author} {\bibinfo {author} {\bibfnamefont {A.}~\bibnamefont
  {Gade}}\ and\ \bibinfo {author} {\bibfnamefont {B.}~\bibnamefont {Sherill}},\
  }\href@noop {} {\bibfield  {journal} {\bibinfo  {journal} {Phys. Scr.}\
  }\textbf {\bibinfo {volume} {91}},\ \bibinfo {pages} {053003} (\bibinfo
  {year} {2016})}\BibitemShut {NoStop}%
\bibitem [{\citenamefont {Morrissey}\ \emph {et~al.}(2003)\citenamefont
  {Morrissey}, \citenamefont {Sherrill}, \citenamefont {Steiner}, \citenamefont
  {Stolz},\ and\ \citenamefont {Wiedenhoever}}]{A1900}%
  \BibitemOpen
  \bibfield  {author} {\bibinfo {author} {\bibfnamefont {D.}~\bibnamefont
  {Morrissey}}, \bibinfo {author} {\bibfnamefont {B.}~\bibnamefont {Sherrill}},
  \bibinfo {author} {\bibfnamefont {M.}~\bibnamefont {Steiner}}, \bibinfo
  {author} {\bibfnamefont {A.}~\bibnamefont {Stolz}},\ and\ \bibinfo {author}
  {\bibfnamefont {I.}~\bibnamefont {Wiedenhoever}},\ }\bibfield  {title}
  {\bibinfo {title} {Commissioning the a1900 projectile fragment separator},\
  }\href {https://doi.org/https://doi.org/10.1016/S0168-583X(02)01895-5}
  {\bibfield  {journal} {\bibinfo  {journal} {Nuclear Instruments and Methods
  in Physics Research Section B: Beam Interactions with Materials and Atoms}\
  }\textbf {\bibinfo {volume} {204}},\ \bibinfo {pages} {90} (\bibinfo {year}
  {2003})},\ \bibinfo {note} {14th International Conference on Electromagnetic
  Isotope Separators and Techniques Related to their Applications}\BibitemShut
  {NoStop}%
\bibitem [{\citenamefont {Prisciandaro}\ \emph {et~al.}(2003)\citenamefont
  {Prisciandaro}, \citenamefont {Morton},\ and\ \citenamefont
  {Mantica}}]{prisci}%
  \BibitemOpen
  \bibfield  {author} {\bibinfo {author} {\bibfnamefont {J.}~\bibnamefont
  {Prisciandaro}}, \bibinfo {author} {\bibfnamefont {A.}~\bibnamefont
  {Morton}},\ and\ \bibinfo {author} {\bibfnamefont {P.}~\bibnamefont
  {Mantica}},\ }\bibfield  {title} {\bibinfo {title} {Beta counting system for
  fast fragmentation beams},\ }\href
  {https://doi.org/https://doi.org/10.1016/S0168-9002(03)01037-4} {\bibfield
  {journal} {\bibinfo  {journal} {Nuclear Instruments and Methods in Physics
  Research Section A: Accelerators, Spectrometers, Detectors and Associated
  Equipment}\ }\textbf {\bibinfo {volume} {505}},\ \bibinfo {pages} {140}
  (\bibinfo {year} {2003})},\ \bibinfo {note} {proceedings of the tenth
  Symposium on Radiation Measurements and Applications}\BibitemShut {NoStop}%
\bibitem [{\citenamefont {Coursey}\ \emph {et~al.}(1982)\citenamefont
  {Coursey}, \citenamefont {Hoppes},\ and\ \citenamefont {Schima}}]{SRM}%
  \BibitemOpen
  \bibfield  {author} {\bibinfo {author} {\bibfnamefont {B.~M.}\ \bibnamefont
  {Coursey}}, \bibinfo {author} {\bibfnamefont {D.~D.}\ \bibnamefont
  {Hoppes}},\ and\ \bibinfo {author} {\bibfnamefont {F.~J.}\ \bibnamefont
  {Schima}},\ }\bibfield  {title} {\bibinfo {title} {Nscl srm source},\
  }\href@noop {} {\bibfield  {journal} {\bibinfo  {journal} {Nucl. Instr. and
  Meth.}\ }\textbf {\bibinfo {volume} {193}} (\bibinfo {year}
  {1982})}\BibitemShut {NoStop}%
\bibitem [{\citenamefont {Prokop}\ \emph {et~al.}(2014)\citenamefont {Prokop},
  \citenamefont {Liddick}, \citenamefont {Abromeit}, \citenamefont {Chemey},
  \citenamefont {Larson}, \citenamefont {Suchyta},\ and\ \citenamefont
  {Tompkins}}]{prokop}%
  \BibitemOpen
  \bibfield  {author} {\bibinfo {author} {\bibfnamefont {C.}~\bibnamefont
  {Prokop}}, \bibinfo {author} {\bibfnamefont {S.}~\bibnamefont {Liddick}},
  \bibinfo {author} {\bibfnamefont {B.}~\bibnamefont {Abromeit}}, \bibinfo
  {author} {\bibfnamefont {A.}~\bibnamefont {Chemey}}, \bibinfo {author}
  {\bibfnamefont {N.}~\bibnamefont {Larson}}, \bibinfo {author} {\bibfnamefont
  {S.}~\bibnamefont {Suchyta}},\ and\ \bibinfo {author} {\bibfnamefont
  {J.}~\bibnamefont {Tompkins}},\ }\bibfield  {title} {\bibinfo {title}
  {Digital data acquisition system implementation at the national
  superconducting cyclotron laboratory},\ }\href
  {https://doi.org/https://doi.org/10.1016/j.nima.2013.12.044} {\bibfield
  {journal} {\bibinfo  {journal} {Nuclear Instruments and Methods in Physics
  Research Section A: Accelerators, Spectrometers, Detectors and Associated
  Equipment}\ }\textbf {\bibinfo {volume} {741}},\ \bibinfo {pages} {163}
  (\bibinfo {year} {2014})}\BibitemShut {NoStop}%
\bibitem [{\citenamefont {Tripathi}\ \emph {et~al.}(2022)\citenamefont
  {Tripathi}, \citenamefont {Bhattacharya}, \citenamefont {Rubino},
  \citenamefont {Benetti}, \citenamefont {Perello}, \citenamefont {Tabor},
  \citenamefont {Liddick}, \citenamefont {Bender}, \citenamefont {Carpenter},
  \citenamefont {Carroll}, \citenamefont {Chester}, \citenamefont {Chiara},
  \citenamefont {Childers}, \citenamefont {Clark}, \citenamefont {Crider},
  \citenamefont {Harke}, \citenamefont {Longfellow}, \citenamefont {Lubna},
  \citenamefont {Luitel}, \citenamefont {Ogunbeku}, \citenamefont {Richard},
  \citenamefont {Saha}, \citenamefont {Shimizu}, \citenamefont {Shehu},
  \citenamefont {Utsuno}, \citenamefont {Unz}, \citenamefont {Xiao},
  \citenamefont {Yoshida},\ and\ \citenamefont
  {Zhu}}]{Vandana_P_S_decay2022PRC}%
  \BibitemOpen
  \bibfield  {author} {\bibinfo {author} {\bibfnamefont {V.}~\bibnamefont
  {Tripathi}}, \bibinfo {author} {\bibfnamefont {S.}~\bibnamefont
  {Bhattacharya}}, \bibinfo {author} {\bibfnamefont {E.}~\bibnamefont
  {Rubino}}, \bibinfo {author} {\bibfnamefont {C.}~\bibnamefont {Benetti}},
  \bibinfo {author} {\bibfnamefont {J.~F.}\ \bibnamefont {Perello}}, \bibinfo
  {author} {\bibfnamefont {S.~L.}\ \bibnamefont {Tabor}}, \bibinfo {author}
  {\bibfnamefont {S.~N.}\ \bibnamefont {Liddick}}, \bibinfo {author}
  {\bibfnamefont {P.~C.}\ \bibnamefont {Bender}}, \bibinfo {author}
  {\bibfnamefont {M.~P.}\ \bibnamefont {Carpenter}}, \bibinfo {author}
  {\bibfnamefont {J.~J.}\ \bibnamefont {Carroll}}, \bibinfo {author}
  {\bibfnamefont {A.}~\bibnamefont {Chester}}, \bibinfo {author} {\bibfnamefont
  {C.~J.}\ \bibnamefont {Chiara}}, \bibinfo {author} {\bibfnamefont
  {K.}~\bibnamefont {Childers}}, \bibinfo {author} {\bibfnamefont {B.~R.}\
  \bibnamefont {Clark}}, \bibinfo {author} {\bibfnamefont {B.~P.}\ \bibnamefont
  {Crider}}, \bibinfo {author} {\bibfnamefont {J.~T.}\ \bibnamefont {Harke}},
  \bibinfo {author} {\bibfnamefont {B.}~\bibnamefont {Longfellow}}, \bibinfo
  {author} {\bibfnamefont {R.~S.}\ \bibnamefont {Lubna}}, \bibinfo {author}
  {\bibfnamefont {S.}~\bibnamefont {Luitel}}, \bibinfo {author} {\bibfnamefont
  {T.~H.}\ \bibnamefont {Ogunbeku}}, \bibinfo {author} {\bibfnamefont {A.~L.}\
  \bibnamefont {Richard}}, \bibinfo {author} {\bibfnamefont {S.}~\bibnamefont
  {Saha}}, \bibinfo {author} {\bibfnamefont {N.}~\bibnamefont {Shimizu}},
  \bibinfo {author} {\bibfnamefont {O.~A.}\ \bibnamefont {Shehu}}, \bibinfo
  {author} {\bibfnamefont {Y.}~\bibnamefont {Utsuno}}, \bibinfo {author}
  {\bibfnamefont {R.}~\bibnamefont {Unz}}, \bibinfo {author} {\bibfnamefont
  {Y.}~\bibnamefont {Xiao}}, \bibinfo {author} {\bibfnamefont {S.}~\bibnamefont
  {Yoshida}},\ and\ \bibinfo {author} {\bibfnamefont {Y.}~\bibnamefont {Zhu}},\
  }\bibfield  {title} {\bibinfo {title}
  {${\ensuremath{\beta}}^{\ensuremath{-}}$ decay of exotic p and s isotopes
  with neutron number near 28},\ }\href
  {https://doi.org/10.1103/PhysRevC.106.064314} {\bibfield  {journal} {\bibinfo
   {journal} {Phys. Rev. C}\ }\textbf {\bibinfo {volume} {106}},\ \bibinfo
  {pages} {064314} (\bibinfo {year} {2022})}\BibitemShut {NoStop}%
\bibitem [{\citenamefont {Sorlin}\ \emph {et~al.}(1993)\citenamefont {Sorlin},
  \citenamefont {Guillemaud-Mueller}, \citenamefont {Mueller}, \citenamefont
  {Borrel}, \citenamefont {Dogny}, \citenamefont {Pougheon}, \citenamefont
  {Kratz}, \citenamefont {Gabelmann}, \citenamefont {Pfeiffer}, \citenamefont
  {W\"ohr}, \citenamefont {Ziegert}, \citenamefont {Penionzhkevich},
  \citenamefont {Lukyanov}, \citenamefont {Salamatin}, \citenamefont {Anne},
  \citenamefont {Borcea}, \citenamefont {Fifield}, \citenamefont {Lewitowicz},
  \citenamefont {Saint-Laurent}, \citenamefont {Bazin}, \citenamefont
  {D\'etraz}, \citenamefont {Thielemann},\ and\ \citenamefont
  {Hillebrandt}}]{Decay_S_Cl_sorlin}%
  \BibitemOpen
  \bibfield  {author} {\bibinfo {author} {\bibfnamefont {O.}~\bibnamefont
  {Sorlin}}, \bibinfo {author} {\bibfnamefont {D.}~\bibnamefont
  {Guillemaud-Mueller}}, \bibinfo {author} {\bibfnamefont {A.~C.}\ \bibnamefont
  {Mueller}}, \bibinfo {author} {\bibfnamefont {V.}~\bibnamefont {Borrel}},
  \bibinfo {author} {\bibfnamefont {S.}~\bibnamefont {Dogny}}, \bibinfo
  {author} {\bibfnamefont {F.}~\bibnamefont {Pougheon}}, \bibinfo {author}
  {\bibfnamefont {K.-L.}\ \bibnamefont {Kratz}}, \bibinfo {author}
  {\bibfnamefont {H.}~\bibnamefont {Gabelmann}}, \bibinfo {author}
  {\bibfnamefont {B.}~\bibnamefont {Pfeiffer}}, \bibinfo {author}
  {\bibfnamefont {A.}~\bibnamefont {W\"ohr}}, \bibinfo {author} {\bibfnamefont
  {W.}~\bibnamefont {Ziegert}}, \bibinfo {author} {\bibfnamefont {Y.~E.}\
  \bibnamefont {Penionzhkevich}}, \bibinfo {author} {\bibfnamefont {S.~M.}\
  \bibnamefont {Lukyanov}}, \bibinfo {author} {\bibfnamefont {V.~S.}\
  \bibnamefont {Salamatin}}, \bibinfo {author} {\bibfnamefont {R.}~\bibnamefont
  {Anne}}, \bibinfo {author} {\bibfnamefont {C.}~\bibnamefont {Borcea}},
  \bibinfo {author} {\bibfnamefont {L.~K.}\ \bibnamefont {Fifield}}, \bibinfo
  {author} {\bibfnamefont {M.}~\bibnamefont {Lewitowicz}}, \bibinfo {author}
  {\bibfnamefont {M.~G.}\ \bibnamefont {Saint-Laurent}}, \bibinfo {author}
  {\bibfnamefont {D.}~\bibnamefont {Bazin}}, \bibinfo {author} {\bibfnamefont
  {C.}~\bibnamefont {D\'etraz}}, \bibinfo {author} {\bibfnamefont {F.-K.}\
  \bibnamefont {Thielemann}},\ and\ \bibinfo {author} {\bibfnamefont
  {W.}~\bibnamefont {Hillebrandt}},\ }\bibfield  {title} {\bibinfo {title}
  {Decay properties of exotic n\ensuremath{\simeq}28 s and cl nuclei and the
  $^{48}\mathrm{Ca}$${/}^{46}$ca abundance ratio},\ }\href
  {https://doi.org/10.1103/PhysRevC.47.2941} {\bibfield  {journal} {\bibinfo
  {journal} {Phys. Rev. C}\ }\textbf {\bibinfo {volume} {47}},\ \bibinfo
  {pages} {2941} (\bibinfo {year} {1993})}\BibitemShut {NoStop}%
\bibitem [{\citenamefont {Mrázek}\ \emph {et~al.}(2004)\citenamefont
  {Mrázek}, \citenamefont {Grévy}, \citenamefont {Iulian}, \citenamefont
  {Buta}, \citenamefont {Negoita}, \citenamefont {Angélique}, \citenamefont
  {Baumann}, \citenamefont {Borcea}, \citenamefont {Canchel}, \citenamefont
  {Catford}, \citenamefont {Courtin}, \citenamefont {Daugas}, \citenamefont
  {Dlouhý}, \citenamefont {Dessagne}, \citenamefont {Knipper}, \citenamefont
  {Lehrsenneau}, \citenamefont {Lecolley}, \citenamefont {Lecouey},
  \citenamefont {Lewitowicz}, \citenamefont {Liénard}, \citenamefont
  {Lukyanov}, \citenamefont {Maréchal}, \citenamefont {Miehe}, \citenamefont
  {{de Oliveira}}, \citenamefont {Orr}, \citenamefont {Pantelica},
  \citenamefont {Penionzhkevich}, \citenamefont {Peter}, \citenamefont
  {Pietri}, \citenamefont {Poirier}, \citenamefont {Sorlin}, \citenamefont
  {Stanoiu}, \citenamefont {Stodel}, \citenamefont {Tarasov},\ and\
  \citenamefont {Timis}}]{Mrazek_44-45_Ar}%
  \BibitemOpen
  \bibfield  {author} {\bibinfo {author} {\bibfnamefont {J.}~\bibnamefont
  {Mrázek}}, \bibinfo {author} {\bibfnamefont {S.}~\bibnamefont {Grévy}},
  \bibinfo {author} {\bibfnamefont {S.}~\bibnamefont {Iulian}}, \bibinfo
  {author} {\bibfnamefont {A.}~\bibnamefont {Buta}}, \bibinfo {author}
  {\bibfnamefont {F.}~\bibnamefont {Negoita}}, \bibinfo {author} {\bibfnamefont
  {J.}~\bibnamefont {Angélique}}, \bibinfo {author} {\bibfnamefont
  {P.}~\bibnamefont {Baumann}}, \bibinfo {author} {\bibfnamefont
  {C.}~\bibnamefont {Borcea}}, \bibinfo {author} {\bibfnamefont
  {G.}~\bibnamefont {Canchel}}, \bibinfo {author} {\bibfnamefont
  {W.}~\bibnamefont {Catford}}, \bibinfo {author} {\bibfnamefont
  {S.}~\bibnamefont {Courtin}}, \bibinfo {author} {\bibfnamefont
  {J.}~\bibnamefont {Daugas}}, \bibinfo {author} {\bibfnamefont
  {Z.}~\bibnamefont {Dlouhý}}, \bibinfo {author} {\bibfnamefont
  {P.}~\bibnamefont {Dessagne}}, \bibinfo {author} {\bibfnamefont
  {A.}~\bibnamefont {Knipper}}, \bibinfo {author} {\bibfnamefont
  {G.}~\bibnamefont {Lehrsenneau}}, \bibinfo {author} {\bibfnamefont
  {F.}~\bibnamefont {Lecolley}}, \bibinfo {author} {\bibfnamefont
  {J.}~\bibnamefont {Lecouey}}, \bibinfo {author} {\bibfnamefont
  {M.}~\bibnamefont {Lewitowicz}}, \bibinfo {author} {\bibfnamefont
  {E.}~\bibnamefont {Liénard}}, \bibinfo {author} {\bibfnamefont
  {S.}~\bibnamefont {Lukyanov}}, \bibinfo {author} {\bibfnamefont
  {F.}~\bibnamefont {Maréchal}}, \bibinfo {author} {\bibfnamefont
  {C.}~\bibnamefont {Miehe}}, \bibinfo {author} {\bibfnamefont
  {F.}~\bibnamefont {{de Oliveira}}}, \bibinfo {author} {\bibfnamefont
  {N.}~\bibnamefont {Orr}}, \bibinfo {author} {\bibfnamefont {D.}~\bibnamefont
  {Pantelica}}, \bibinfo {author} {\bibfnamefont {Y.}~\bibnamefont
  {Penionzhkevich}}, \bibinfo {author} {\bibfnamefont {J.}~\bibnamefont
  {Peter}}, \bibinfo {author} {\bibfnamefont {S.}~\bibnamefont {Pietri}},
  \bibinfo {author} {\bibfnamefont {E.}~\bibnamefont {Poirier}}, \bibinfo
  {author} {\bibfnamefont {O.}~\bibnamefont {Sorlin}}, \bibinfo {author}
  {\bibfnamefont {M.}~\bibnamefont {Stanoiu}}, \bibinfo {author} {\bibfnamefont
  {O.}~\bibnamefont {Stodel}}, \bibinfo {author} {\bibfnamefont
  {O.}~\bibnamefont {Tarasov}},\ and\ \bibinfo {author} {\bibfnamefont
  {C.}~\bibnamefont {Timis}},\ }\bibfield  {title} {\bibinfo {title} {Study of
  neutron-rich argon isotopes in $\beta$-decay},\ }\href
  {https://doi.org/https://doi.org/10.1016/j.nuclphysa.2004.03.021} {\bibfield
  {journal} {\bibinfo  {journal} {Nuclear Physics A}\ }\textbf {\bibinfo
  {volume} {734}},\ \bibinfo {pages} {E65} (\bibinfo {year} {2004})},\ \bibinfo
  {note} {proceedings of the Eighth International Conference on Nucleus-Nucleus
  Collisions (NN2003)}\BibitemShut {NoStop}%
\bibitem [{\citenamefont {Lubna}\ \emph {et~al.}(2020)\citenamefont {Lubna},
  \citenamefont {Kravvaris}, \citenamefont {Tabor}, \citenamefont {Tripathi},
  \citenamefont {Rubino},\ and\ \citenamefont {Volya}}]{Rebeka_SM}%
  \BibitemOpen
  \bibfield  {author} {\bibinfo {author} {\bibfnamefont {R.~S.}\ \bibnamefont
  {Lubna}}, \bibinfo {author} {\bibfnamefont {K.}~\bibnamefont {Kravvaris}},
  \bibinfo {author} {\bibfnamefont {S.~L.}\ \bibnamefont {Tabor}}, \bibinfo
  {author} {\bibfnamefont {V.}~\bibnamefont {Tripathi}}, \bibinfo {author}
  {\bibfnamefont {E.}~\bibnamefont {Rubino}},\ and\ \bibinfo {author}
  {\bibfnamefont {A.}~\bibnamefont {Volya}},\ }\bibfield  {title} {\bibinfo
  {title} {Evolution of the $n=20$ and 28 shell gaps and two-particle-two-hole
  states in the fsu interaction},\ }\href
  {https://doi.org/10.1103/PhysRevResearch.2.043342} {\bibfield  {journal}
  {\bibinfo  {journal} {Phys. Rev. Res.}\ }\textbf {\bibinfo {volume} {2}},\
  \bibinfo {pages} {043342} (\bibinfo {year} {2020})}\BibitemShut {NoStop}%
\bibitem [{\citenamefont {Wan}\ \emph {et~al.}(1999)\citenamefont {Wan},
  \citenamefont {Gerl}, \citenamefont {Cub}, \citenamefont {Holeczek},
  \citenamefont {Reiter}, \citenamefont {Schwalm}, \citenamefont {Aumann},
  \citenamefont {Boretzky}, \citenamefont {Dostal}, \citenamefont {Eberlein},
  \citenamefont {Emling}, \citenamefont {Ender}, \citenamefont {Elze},
  \citenamefont {Geissel}, \citenamefont {Grunschlob}, \citenamefont
  {Holzmann}, \citenamefont {Iwasa}, \citenamefont {Kaspar}, \citenamefont
  {Kleinbohl}, \citenamefont {Koschorrek}, \citenamefont {Leifels},
  \citenamefont {Leistenschneider}, \citenamefont {Peter}, \citenamefont
  {Schaffner}, \citenamefont {Scheidenberger}, \citenamefont {Schubart},
  \citenamefont {Schubert}, \citenamefont {Simon}, \citenamefont {Stengel},
  \citenamefont {Surowiec},\ and\ \citenamefont {Wollersheim}}]{Wan_44Ar_deep}%
  \BibitemOpen
  \bibfield  {author} {\bibinfo {author} {\bibfnamefont {S.}~\bibnamefont
  {Wan}}, \bibinfo {author} {\bibfnamefont {J.}~\bibnamefont {Gerl}}, \bibinfo
  {author} {\bibfnamefont {J.}~\bibnamefont {Cub}}, \bibinfo {author}
  {\bibfnamefont {J.}~\bibnamefont {Holeczek}}, \bibinfo {author}
  {\bibfnamefont {P.}~\bibnamefont {Reiter}}, \bibinfo {author} {\bibfnamefont
  {D.}~\bibnamefont {Schwalm}}, \bibinfo {author} {\bibfnamefont
  {T.}~\bibnamefont {Aumann}}, \bibinfo {author} {\bibfnamefont
  {K.}~\bibnamefont {Boretzky}}, \bibinfo {author} {\bibfnamefont
  {W.}~\bibnamefont {Dostal}}, \bibinfo {author} {\bibfnamefont
  {B.}~\bibnamefont {Eberlein}}, \bibinfo {author} {\bibfnamefont
  {H.}~\bibnamefont {Emling}}, \bibinfo {author} {\bibfnamefont
  {C.}~\bibnamefont {Ender}}, \bibinfo {author} {\bibfnamefont
  {T.}~\bibnamefont {Elze}}, \bibinfo {author} {\bibfnamefont {H.}~\bibnamefont
  {Geissel}}, \bibinfo {author} {\bibfnamefont {A.}~\bibnamefont {Grunschlob}},
  \bibinfo {author} {\bibfnamefont {R.}~\bibnamefont {Holzmann}}, \bibinfo
  {author} {\bibfnamefont {N.}~\bibnamefont {Iwasa}}, \bibinfo {author}
  {\bibfnamefont {M.}~\bibnamefont {Kaspar}}, \bibinfo {author} {\bibfnamefont
  {A.}~\bibnamefont {Kleinbohl}}, \bibinfo {author} {\bibfnamefont
  {O.}~\bibnamefont {Koschorrek}}, \bibinfo {author} {\bibfnamefont
  {Y.}~\bibnamefont {Leifels}}, \bibinfo {author} {\bibfnamefont
  {A.}~\bibnamefont {Leistenschneider}}, \bibinfo {author} {\bibfnamefont
  {I.}~\bibnamefont {Peter}}, \bibinfo {author} {\bibfnamefont
  {H.}~\bibnamefont {Schaffner}}, \bibinfo {author} {\bibfnamefont
  {C.}~\bibnamefont {Scheidenberger}}, \bibinfo {author} {\bibfnamefont
  {R.}~\bibnamefont {Schubart}}, \bibinfo {author} {\bibfnamefont
  {R.}~\bibnamefont {Schubert}}, \bibinfo {author} {\bibfnamefont
  {H.}~\bibnamefont {Simon}}, \bibinfo {author} {\bibfnamefont
  {G.}~\bibnamefont {Stengel}}, \bibinfo {author} {\bibfnamefont
  {A.}~\bibnamefont {Surowiec}},\ and\ \bibinfo {author} {\bibfnamefont
  {H.}~\bibnamefont {Wollersheim}},\ }\bibfield  {title} {\bibinfo {title}
  {In-beam $\gamma$-spectroscopy with relativistic radioactive ion beams},\
  }\href {https://doi.org/https://doi.org/10.1007/s100500050331} {\bibfield
  {journal} {\bibinfo  {journal} {European Physics Journal A}\ }\textbf
  {\bibinfo {volume} {6}},\ \bibinfo {pages} {167} (\bibinfo {year}
  {1999})}\BibitemShut {NoStop}%
\bibitem [{\citenamefont {{B. Fornal}}\ \emph {et~al.}(2000)\citenamefont {{B.
  Fornal}}, \citenamefont {{R. Broda}}, \citenamefont {{W. Kr\'olas}},
  \citenamefont {{T. Pawlat}}, \citenamefont {{J. Wrzesi\'{}nski}},
  \citenamefont {{D. Bazzacco}}, \citenamefont {{S. Lunardi}}, \citenamefont
  {{C. Rossi Alvarez}}, \citenamefont {{G. Viesti}}, \citenamefont {{G. de
  Angelis}}, \citenamefont {{M. Cinausero}}, \citenamefont {{D. Napoli}},
  \citenamefont {{J. Gerl}}, \citenamefont {{E. Caurier}},\ and\ \citenamefont
  {{F. Nowacki}}}]{44Ar_deep_Fornal}%
  \BibitemOpen
  \bibfield  {author} {\bibinfo {author} {\bibnamefont {{B. Fornal}}}, \bibinfo
  {author} {\bibnamefont {{R. Broda}}}, \bibinfo {author} {\bibnamefont {{W.
  Kr\'olas}}}, \bibinfo {author} {\bibnamefont {{T. Pawlat}}}, \bibinfo
  {author} {\bibnamefont {{J. Wrzesi\'{}nski}}}, \bibinfo {author}
  {\bibnamefont {{D. Bazzacco}}}, \bibinfo {author} {\bibnamefont {{S.
  Lunardi}}}, \bibinfo {author} {\bibnamefont {{C. Rossi Alvarez}}}, \bibinfo
  {author} {\bibnamefont {{G. Viesti}}}, \bibinfo {author} {\bibnamefont {{G.
  de Angelis}}}, \bibinfo {author} {\bibnamefont {{M. Cinausero}}}, \bibinfo
  {author} {\bibnamefont {{D. Napoli}}}, \bibinfo {author} {\bibnamefont {{J.
  Gerl}}}, \bibinfo {author} {\bibnamefont {{E. Caurier}}},\ and\ \bibinfo
  {author} {\bibnamefont {{F. Nowacki}}},\ }\bibfield  {title} {\bibinfo
  {title} {New states in $^{44,46}\mathrm{Ar}$ isotopes from deep-inelastic
  heavy ion reaction studies},\ }\href {https://doi.org/epja/v7/p147(epja172)}
  {\bibfield  {journal} {\bibinfo  {journal} {Eur. Phys. J. A}\ }\textbf
  {\bibinfo {volume} {7}},\ \bibinfo {pages} {147} (\bibinfo {year}
  {2000})}\BibitemShut {NoStop}%
\bibitem [{\citenamefont {Volya}(2016)}]{volya}%
  \BibitemOpen
  \bibfield  {author} {\bibinfo {author} {\bibfnamefont {A.}~\bibnamefont
  {Volya}},\ }\href {https://www.volya.net/.} {\bibinfo {title} {Continuum
  shell model code,https://www.volya.net/.}} (\bibinfo {year}
  {2016})\BibitemShut {NoStop}%
\bibitem [{\citenamefont {Gade}\ \emph {et~al.}(2022)\citenamefont {Gade},
  \citenamefont {Brown}, \citenamefont {Weisshaar}, \citenamefont {Bazin},
  \citenamefont {Brown}, \citenamefont {Charity}, \citenamefont {Farris},
  \citenamefont {Hill}, \citenamefont {Li}, \citenamefont {Longfellow},
  \citenamefont {Rhodes}, \citenamefont {Reviol},\ and\ \citenamefont
  {Tostevin}}]{Gade2022}%
  \BibitemOpen
  \bibfield  {author} {\bibinfo {author} {\bibfnamefont {A.}~\bibnamefont
  {Gade}}, \bibinfo {author} {\bibfnamefont {B.~A.}\ \bibnamefont {Brown}},
  \bibinfo {author} {\bibfnamefont {D.}~\bibnamefont {Weisshaar}}, \bibinfo
  {author} {\bibfnamefont {D.}~\bibnamefont {Bazin}}, \bibinfo {author}
  {\bibfnamefont {K.~W.}\ \bibnamefont {Brown}}, \bibinfo {author}
  {\bibfnamefont {R.~J.}\ \bibnamefont {Charity}}, \bibinfo {author}
  {\bibfnamefont {P.}~\bibnamefont {Farris}}, \bibinfo {author} {\bibfnamefont
  {A.~M.}\ \bibnamefont {Hill}}, \bibinfo {author} {\bibfnamefont
  {J.}~\bibnamefont {Li}}, \bibinfo {author} {\bibfnamefont {B.}~\bibnamefont
  {Longfellow}}, \bibinfo {author} {\bibfnamefont {D.}~\bibnamefont {Rhodes}},
  \bibinfo {author} {\bibfnamefont {W.}~\bibnamefont {Reviol}},\ and\ \bibinfo
  {author} {\bibfnamefont {J.~A.}\ \bibnamefont {Tostevin}},\ }\bibfield
  {title} {\bibinfo {title} {Dissipative reactions with intermediate-energy
  beams: A novel approach to populate complex-structure states in rare
  isotopes},\ }\href {https://doi.org/10.1103/PhysRevLett.129.242501}
  {\bibfield  {journal} {\bibinfo  {journal} {Phys. Rev. Lett.}\ }\textbf
  {\bibinfo {volume} {129}},\ \bibinfo {pages} {242501} (\bibinfo {year}
  {2022})}\BibitemShut {NoStop}%
\bibitem [{\citenamefont {Brown}(2022)}]{Brown2022}%
  \BibitemOpen
  \bibfield  {author} {\bibinfo {author} {\bibfnamefont {B. A.}\ \bibnamefont
  {Brown}},\ }\bibfield  {title} {\bibinfo {title} {The nuclear shell model
  towards the drip lines},\ }\href {https://doi.org/10.3390/physics4020035}
  {\bibfield  {journal} {\bibinfo  {journal} {Physics}\ }\textbf {\bibinfo
  {volume} {4}},\ \bibinfo {pages} {525} (\bibinfo {year} {2022})}\BibitemShut
  {NoStop}%
\bibitem [{\citenamefont {Gaudefroy}\ \emph {et~al.}(2005)\citenamefont
  {Gaudefroy}, \citenamefont {Sorlin}, \citenamefont {Beaumel}, \citenamefont
  {Blumenfeld}, \citenamefont {Dombrádi}, \citenamefont {Fortier},
  \citenamefont {Franchoo}, \citenamefont {Gélin}, \citenamefont {Gibelin},
  \citenamefont {Grévy}, \citenamefont {Hammache}, \citenamefont {Ibrahim},
  \citenamefont {Kemper}, \citenamefont {Kratz}, \citenamefont {Lukyanov},
  \citenamefont {Monrozeau}, \citenamefont {Nalpas}, \citenamefont {Nowacki},
  \citenamefont {Ostrowski}, \citenamefont {Penionzhkevich}, \citenamefont
  {Pollaco}, \citenamefont {Roussel-Chomaz}, \citenamefont {Rich},
  \citenamefont {Scarpaci}, \citenamefont {Laurent}, \citenamefont {Sohler},
  \citenamefont {Stanoiu}, \citenamefont {Tryggestadt},\ and\ \citenamefont
  {Verney}}]{Gaudefroy_2005}%
  \BibitemOpen
  \bibfield  {author} {\bibinfo {author} {\bibfnamefont {L.}~\bibnamefont
  {Gaudefroy}}, \bibinfo {author} {\bibfnamefont {O.}~\bibnamefont {Sorlin}},
  \bibinfo {author} {\bibfnamefont {D.}~\bibnamefont {Beaumel}}, \bibinfo
  {author} {\bibfnamefont {Y.}~\bibnamefont {Blumenfeld}}, \bibinfo {author}
  {\bibfnamefont {Z.}~\bibnamefont {Dombrádi}}, \bibinfo {author}
  {\bibfnamefont {S.}~\bibnamefont {Fortier}}, \bibinfo {author} {\bibfnamefont
  {S.}~\bibnamefont {Franchoo}}, \bibinfo {author} {\bibfnamefont
  {M.}~\bibnamefont {Gélin}}, \bibinfo {author} {\bibfnamefont
  {J.}~\bibnamefont {Gibelin}}, \bibinfo {author} {\bibfnamefont
  {S.}~\bibnamefont {Grévy}}, \bibinfo {author} {\bibfnamefont
  {F.}~\bibnamefont {Hammache}}, \bibinfo {author} {\bibfnamefont
  {F.}~\bibnamefont {Ibrahim}}, \bibinfo {author} {\bibfnamefont
  {K.}~\bibnamefont {Kemper}}, \bibinfo {author} {\bibfnamefont {K.~L.}\
  \bibnamefont {Kratz}}, \bibinfo {author} {\bibfnamefont {S.~M.}\ \bibnamefont
  {Lukyanov}}, \bibinfo {author} {\bibfnamefont {C.}~\bibnamefont {Monrozeau}},
  \bibinfo {author} {\bibfnamefont {L.}~\bibnamefont {Nalpas}}, \bibinfo
  {author} {\bibfnamefont {F.}~\bibnamefont {Nowacki}}, \bibinfo {author}
  {\bibfnamefont {A.~N.}\ \bibnamefont {Ostrowski}}, \bibinfo {author}
  {\bibfnamefont {Y.-E.}\ \bibnamefont {Penionzhkevich}}, \bibinfo {author}
  {\bibfnamefont {E.}~\bibnamefont {Pollaco}}, \bibinfo {author} {\bibfnamefont
  {P.}~\bibnamefont {Roussel-Chomaz}}, \bibinfo {author} {\bibfnamefont
  {E.}~\bibnamefont {Rich}}, \bibinfo {author} {\bibfnamefont {J.~A.}\
  \bibnamefont {Scarpaci}}, \bibinfo {author} {\bibfnamefont {M.~G.~S.}\
  \bibnamefont {Laurent}}, \bibinfo {author} {\bibfnamefont {D.}~\bibnamefont
  {Sohler}}, \bibinfo {author} {\bibfnamefont {M.}~\bibnamefont {Stanoiu}},
  \bibinfo {author} {\bibfnamefont {E.}~\bibnamefont {Tryggestadt}},\ and\
  \bibinfo {author} {\bibfnamefont {D.}~\bibnamefont {Verney}},\ }\bibfield
  {title} {\bibinfo {title} {Study of 45ar through (d, p) reaction at spiral},\
  }\href {https://doi.org/10.1088/0954-3899/31/10/044} {\bibfield  {journal}
  {\bibinfo  {journal} {Journal of Physics G: Nuclear and Particle Physics}\
  }\textbf {\bibinfo {volume} {31}},\ \bibinfo {pages} {S1623} (\bibinfo {year}
  {2005})}\BibitemShut {NoStop}%
\bibitem [{\citenamefont {Gaudefroy}\ \emph {et~al.}(2008)\citenamefont
  {Gaudefroy}, \citenamefont {Sorlin}, \citenamefont {Nowacki}, \citenamefont
  {Beaumel}, \citenamefont {Blumenfeld}, \citenamefont {Dombr\'adi},
  \citenamefont {Fortier}, \citenamefont {Franchoo}, \citenamefont {Gr\'evy},
  \citenamefont {Hammache}, \citenamefont {Kemper}, \citenamefont {Kratz},
  \citenamefont {St.~Laurent}, \citenamefont {Lukyanov}, \citenamefont
  {Nalpas}, \citenamefont {Ostrowski}, \citenamefont {Penionzhkevich},
  \citenamefont {Pollacco}, \citenamefont {Roussel}, \citenamefont
  {Roussel-Chomaz}, \citenamefont {Sohler}, \citenamefont {Stanoiu},\ and\
  \citenamefont {Tryggestad}}]{45Ar_dp_2008}%
  \BibitemOpen
  \bibfield  {author} {\bibinfo {author} {\bibfnamefont {L.}~\bibnamefont
  {Gaudefroy}}, \bibinfo {author} {\bibfnamefont {O.}~\bibnamefont {Sorlin}},
  \bibinfo {author} {\bibfnamefont {F.}~\bibnamefont {Nowacki}}, \bibinfo
  {author} {\bibfnamefont {D.}~\bibnamefont {Beaumel}}, \bibinfo {author}
  {\bibfnamefont {Y.}~\bibnamefont {Blumenfeld}}, \bibinfo {author}
  {\bibfnamefont {Z.}~\bibnamefont {Dombr\'adi}}, \bibinfo {author}
  {\bibfnamefont {S.}~\bibnamefont {Fortier}}, \bibinfo {author} {\bibfnamefont
  {S.}~\bibnamefont {Franchoo}}, \bibinfo {author} {\bibfnamefont
  {S.}~\bibnamefont {Gr\'evy}}, \bibinfo {author} {\bibfnamefont
  {F.}~\bibnamefont {Hammache}}, \bibinfo {author} {\bibfnamefont {K.~W.}\
  \bibnamefont {Kemper}}, \bibinfo {author} {\bibfnamefont {K.~L.}\
  \bibnamefont {Kratz}}, \bibinfo {author} {\bibfnamefont {M.~G.}\ \bibnamefont
  {St.~Laurent}}, \bibinfo {author} {\bibfnamefont {S.~M.}\ \bibnamefont
  {Lukyanov}}, \bibinfo {author} {\bibfnamefont {L.}~\bibnamefont {Nalpas}},
  \bibinfo {author} {\bibfnamefont {A.~N.}\ \bibnamefont {Ostrowski}}, \bibinfo
  {author} {\bibfnamefont {Y.-E.}\ \bibnamefont {Penionzhkevich}}, \bibinfo
  {author} {\bibfnamefont {E.~C.}\ \bibnamefont {Pollacco}}, \bibinfo {author}
  {\bibfnamefont {P.}~\bibnamefont {Roussel}}, \bibinfo {author} {\bibfnamefont
  {P.}~\bibnamefont {Roussel-Chomaz}}, \bibinfo {author} {\bibfnamefont
  {D.}~\bibnamefont {Sohler}}, \bibinfo {author} {\bibfnamefont
  {M.}~\bibnamefont {Stanoiu}},\ and\ \bibinfo {author} {\bibfnamefont
  {E.}~\bibnamefont {Tryggestad}},\ }\bibfield  {title} {\bibinfo {title}
  {Structure of the $n=27$ isotones derived from the
  $^{44}\mathrm{Ar}$($d,p$)$^{45}\mathrm{Ar}$ reaction},\ }\href
  {https://doi.org/10.1103/PhysRevC.78.034307} {\bibfield  {journal} {\bibinfo
  {journal} {Phys. Rev. C}\ }\textbf {\bibinfo {volume} {78}},\ \bibinfo
  {pages} {034307} (\bibinfo {year} {2008})}\BibitemShut {NoStop}%
\bibitem [{\citenamefont {Szilner}\ \emph {et~al.}(2011)\citenamefont
  {Szilner}, \citenamefont {Corradi}, \citenamefont {Haas}, \citenamefont
  {Lebhertz}, \citenamefont {Pollarolo}, \citenamefont {Ur}, \citenamefont
  {Angus}, \citenamefont {Beghini}, \citenamefont {Bouhelal}, \citenamefont
  {Chapman}, \citenamefont {Caurier}, \citenamefont {Courtin}, \citenamefont
  {Farnea}, \citenamefont {Fioretto}, \citenamefont {Gadea}, \citenamefont
  {Goasduff}, \citenamefont {Jelavi\ifmmode~\acute{c}\else \'{c}\fi{}-Malenica}, \citenamefont
  {Kumar}, \citenamefont {Lunardi}, \citenamefont
  {M\ifmmode~\u{a}\else \u{a}\fi{}rginean}, \citenamefont {Mason},
  \citenamefont {Mengoni}, \citenamefont {Montagnoli}, \citenamefont {Nowacki},
  \citenamefont {Recchia}, \citenamefont {Sahin}, \citenamefont {Salsac},
  \citenamefont {Scarlassara}, \citenamefont {Silvestri}, \citenamefont
  {Smith}, \citenamefont {Soi\ifmmode~\acute{c}\else \'{c}\fi{}}, \citenamefont
  {Stefanini},\ and\ \citenamefont {Valiente-Dob\'on}}]{Ar_Szilner_transfer}%
  \BibitemOpen
  \bibfield  {author} {\bibinfo {author} {\bibfnamefont {S.}~\bibnamefont
  {Szilner}}, \bibinfo {author} {\bibfnamefont {L.}~\bibnamefont {Corradi}},
  \bibinfo {author} {\bibfnamefont {F.}~\bibnamefont {Haas}}, \bibinfo {author}
  {\bibfnamefont {D.}~\bibnamefont {Lebhertz}}, \bibinfo {author}
  {\bibfnamefont {G.}~\bibnamefont {Pollarolo}}, \bibinfo {author}
  {\bibfnamefont {C.~A.}\ \bibnamefont {Ur}}, \bibinfo {author} {\bibfnamefont
  {L.}~\bibnamefont {Angus}}, \bibinfo {author} {\bibfnamefont
  {S.}~\bibnamefont {Beghini}}, \bibinfo {author} {\bibfnamefont
  {M.}~\bibnamefont {Bouhelal}}, \bibinfo {author} {\bibfnamefont
  {R.}~\bibnamefont {Chapman}}, \bibinfo {author} {\bibfnamefont
  {E.}~\bibnamefont {Caurier}}, \bibinfo {author} {\bibfnamefont
  {S.}~\bibnamefont {Courtin}}, \bibinfo {author} {\bibfnamefont
  {E.}~\bibnamefont {Farnea}}, \bibinfo {author} {\bibfnamefont
  {E.}~\bibnamefont {Fioretto}}, \bibinfo {author} {\bibfnamefont
  {A.}~\bibnamefont {Gadea}}, \bibinfo {author} {\bibfnamefont
  {A.}~\bibnamefont {Goasduff}}, \bibinfo {author} {\bibfnamefont
  {D.}~\bibnamefont {Jelavi\ifmmode~\acute{c}\else \'{c}\fi{}-Malenica}},
  \bibinfo {author} {\bibfnamefont {V.}~\bibnamefont {Kumar}}, \bibinfo
  {author} {\bibfnamefont {S.}~\bibnamefont {Lunardi}}, \bibinfo {author}
  {\bibfnamefont {N.}~\bibnamefont {M\ifmmode~\u{a}\else \u{a}\fi{}rginean}},
  \bibinfo {author} {\bibfnamefont {P.}~\bibnamefont {Mason}}, \bibinfo
  {author} {\bibfnamefont {D.}~\bibnamefont {Mengoni}}, \bibinfo {author}
  {\bibfnamefont {G.}~\bibnamefont {Montagnoli}}, \bibinfo {author}
  {\bibfnamefont {F.}~\bibnamefont {Nowacki}}, \bibinfo {author} {\bibfnamefont
  {F.}~\bibnamefont {Recchia}}, \bibinfo {author} {\bibfnamefont
  {E.}~\bibnamefont {Sahin}}, \bibinfo {author} {\bibfnamefont {M.-D.}\
  \bibnamefont {Salsac}}, \bibinfo {author} {\bibfnamefont {F.}~\bibnamefont
  {Scarlassara}}, \bibinfo {author} {\bibfnamefont {R.}~\bibnamefont
  {Silvestri}}, \bibinfo {author} {\bibfnamefont {J.~F.}\ \bibnamefont
  {Smith}}, \bibinfo {author} {\bibfnamefont {N.}~\bibnamefont
  {Soi\ifmmode~\acute{c}\else \'{c}\fi{}}}, \bibinfo {author} {\bibfnamefont
  {A.~M.}\ \bibnamefont {Stefanini}},\ and\ \bibinfo {author} {\bibfnamefont
  {J.~J.}\ \bibnamefont {Valiente-Dob\'on}},\ }\bibfield  {title} {\bibinfo
  {title} {Interplay between single-particle and collective excitations in
  argon isotopes populated by transfer reactions},\ }\href
  {https://doi.org/10.1103/PhysRevC.84.014325} {\bibfield  {journal} {\bibinfo
  {journal} {Phys. Rev. C}\ }\textbf {\bibinfo {volume} {84}},\ \bibinfo
  {pages} {014325} (\bibinfo {year} {2011})}\BibitemShut {NoStop}%
\bibitem [{\citenamefont {Mar\'echal}\ \emph {et~al.}(1999)\citenamefont
  {Mar\'echal}, \citenamefont {Suomij\"arvi}, \citenamefont {Blumenfeld},
  \citenamefont {Azhari}, \citenamefont {Bazin}, \citenamefont {Brown},
  \citenamefont {Cottle}, \citenamefont {Fauerbach}, \citenamefont
  {Glasmacher}, \citenamefont {Hirzebruch}, \citenamefont {Jewell},
  \citenamefont {Kelley}, \citenamefont {Kemper}, \citenamefont {Mantica},
  \citenamefont {Morrissey}, \citenamefont {Riley}, \citenamefont {Scarpaci},
  \citenamefont {Scheit},\ and\ \citenamefont {Steiner}}]{43ArProtonScat}%
  \BibitemOpen
  \bibfield  {author} {\bibinfo {author} {\bibfnamefont {F.}~\bibnamefont
  {Mar\'echal}}, \bibinfo {author} {\bibfnamefont {T.}~\bibnamefont
  {Suomij\"arvi}}, \bibinfo {author} {\bibfnamefont {Y.}~\bibnamefont
  {Blumenfeld}}, \bibinfo {author} {\bibfnamefont {A.}~\bibnamefont {Azhari}},
  \bibinfo {author} {\bibfnamefont {D.}~\bibnamefont {Bazin}}, \bibinfo
  {author} {\bibfnamefont {J.~A.}\ \bibnamefont {Brown}}, \bibinfo {author}
  {\bibfnamefont {P.~D.}\ \bibnamefont {Cottle}}, \bibinfo {author}
  {\bibfnamefont {M.}~\bibnamefont {Fauerbach}}, \bibinfo {author}
  {\bibfnamefont {T.}~\bibnamefont {Glasmacher}}, \bibinfo {author}
  {\bibfnamefont {S.~E.}\ \bibnamefont {Hirzebruch}}, \bibinfo {author}
  {\bibfnamefont {J.~K.}\ \bibnamefont {Jewell}}, \bibinfo {author}
  {\bibfnamefont {J.~H.}\ \bibnamefont {Kelley}}, \bibinfo {author}
  {\bibfnamefont {K.~W.}\ \bibnamefont {Kemper}}, \bibinfo {author}
  {\bibfnamefont {P.~F.}\ \bibnamefont {Mantica}}, \bibinfo {author}
  {\bibfnamefont {D.~J.}\ \bibnamefont {Morrissey}}, \bibinfo {author}
  {\bibfnamefont {L.~A.}\ \bibnamefont {Riley}}, \bibinfo {author}
  {\bibfnamefont {J.~A.}\ \bibnamefont {Scarpaci}}, \bibinfo {author}
  {\bibfnamefont {H.}~\bibnamefont {Scheit}},\ and\ \bibinfo {author}
  {\bibfnamefont {M.}~\bibnamefont {Steiner}},\ }\bibfield  {title} {\bibinfo
  {title} {Proton scattering from the unstable neutron-rich nucleus
  ${}^{43}\mathrm{Ar}$},\ }\href {https://doi.org/10.1103/PhysRevC.60.064623}
  {\bibfield  {journal} {\bibinfo  {journal} {Phys. Rev. C}\ }\textbf {\bibinfo
  {volume} {60}},\ \bibinfo {pages} {064623} (\bibinfo {year}
  {1999})}\BibitemShut {NoStop}%
\bibitem [{\citenamefont {Winger}\ \emph {et~al.}(2006)\citenamefont {Winger},
  \citenamefont {Mantica},\ and\ \citenamefont {Ronningen}}]{43Ar_Winger2006}%
  \BibitemOpen
  \bibfield  {author} {\bibinfo {author} {\bibfnamefont {J.~A.}\ \bibnamefont
  {Winger}}, \bibinfo {author} {\bibfnamefont {P.~F.}\ \bibnamefont
  {Mantica}},\ and\ \bibinfo {author} {\bibfnamefont {R.~M.}\ \bibnamefont
  {Ronningen}},\ }\bibfield  {title} {\bibinfo {title} {\ensuremath{\beta}
  decay of $^{40,42}\mathrm{S}$ and $^{43}\mathrm{Cl}$},\ }\href
  {https://doi.org/10.1103/PhysRevC.73.044318} {\bibfield  {journal} {\bibinfo
  {journal} {Phys. Rev. C}\ }\textbf {\bibinfo {volume} {73}},\ \bibinfo
  {pages} {044318} (\bibinfo {year} {2006})}\BibitemShut {NoStop}%
\bibitem [{\citenamefont {Jun}\ \emph {et~al.}(2001)\citenamefont {Jun},
  \citenamefont {Singh},\ and\ \citenamefont {Cameron}}]{nndc44Ar}%
  \BibitemOpen
  \bibfield  {author} {\bibinfo {author} {\bibfnamefont {C.}~\bibnamefont
  {Jun}}, \bibinfo {author} {\bibfnamefont {B.}~\bibnamefont {Singh}},\ and\
  \bibinfo {author} {\bibfnamefont {J.~A.}\ \bibnamefont {Cameron}},\ }\href
  {https://www.nndc.bnl.gov/ensdf/EnsdfDispatcherServlet} {\bibinfo {title}
  {https://www.nndc.bnl.gov/nudat3/}} (\bibinfo {year} {2001})\BibitemShut
  {NoStop}%
\bibitem [{\citenamefont {NNDC}(2005)}]{nndc}%
  \BibitemOpen
  \bibfield  {author} {\bibinfo {author} {\bibnamefont {NNDC}},\ }\href
  {https://www.nndc.bnl.gov/nudat3} {\bibinfo {title}
  {https://www.nndc.bnl.gov/nudat3/}} (\bibinfo {year} {2005})\BibitemShut
  {NoStop}%
\bibitem [{\citenamefont {Speidel}\ \emph {et~al.}(2006)\citenamefont
  {Speidel}, \citenamefont {Schielke}, \citenamefont {Leske}, \citenamefont
  {Gerber}, \citenamefont {Maier-Komor}, \citenamefont {Robinson},
  \citenamefont {Sharon},\ and\ \citenamefont {Zamick}}]{38Ar_Speidel_PLB}%
  \BibitemOpen
  \bibfield  {author} {\bibinfo {author} {\bibfnamefont {K.-H.}\ \bibnamefont
  {Speidel}}, \bibinfo {author} {\bibfnamefont {S.}~\bibnamefont {Schielke}},
  \bibinfo {author} {\bibfnamefont {J.}~\bibnamefont {Leske}}, \bibinfo
  {author} {\bibfnamefont {J.}~\bibnamefont {Gerber}}, \bibinfo {author}
  {\bibfnamefont {P.}~\bibnamefont {Maier-Komor}}, \bibinfo {author}
  {\bibfnamefont {S.}~\bibnamefont {Robinson}}, \bibinfo {author}
  {\bibfnamefont {Y.}~\bibnamefont {Sharon}},\ and\ \bibinfo {author}
  {\bibfnamefont {L.}~\bibnamefont {Zamick}},\ }\bibfield  {title} {\bibinfo
  {title} {Experimental g factors and b(e2) values in ar isotopes: Crossing the
  n=20 semi-magic divide},\ }\href
  {https://doi.org/https://doi.org/10.1016/j.physletb.2005.10.052} {\bibfield
  {journal} {\bibinfo  {journal} {Physics Letters B}\ }\textbf {\bibinfo
  {volume} {632}},\ \bibinfo {pages} {207} (\bibinfo {year}
  {2006})}\BibitemShut {NoStop}%
\bibitem [{\citenamefont {Vanhoy}\ \emph {et~al.}(1992)\citenamefont {Vanhoy},
  \citenamefont {McEllistrem}, \citenamefont {Hicks}, \citenamefont {Gatenby},
  \citenamefont {Baum}, \citenamefont {Johnson}, \citenamefont {Moln\'ar},\
  and\ \citenamefont {Yates}}]{48Ca_Vanhoy1992}%
  \BibitemOpen
  \bibfield  {author} {\bibinfo {author} {\bibfnamefont {J.~R.}\ \bibnamefont
  {Vanhoy}}, \bibinfo {author} {\bibfnamefont {M.~T.}\ \bibnamefont
  {McEllistrem}}, \bibinfo {author} {\bibfnamefont {S.~F.}\ \bibnamefont
  {Hicks}}, \bibinfo {author} {\bibfnamefont {R.~A.}\ \bibnamefont {Gatenby}},
  \bibinfo {author} {\bibfnamefont {E.~M.}\ \bibnamefont {Baum}}, \bibinfo
  {author} {\bibfnamefont {E.~L.}\ \bibnamefont {Johnson}}, \bibinfo {author}
  {\bibfnamefont {G.}~\bibnamefont {Moln\'ar}},\ and\ \bibinfo {author}
  {\bibfnamefont {S.~W.}\ \bibnamefont {Yates}},\ }\bibfield  {title} {\bibinfo
  {title} {Neutron and proton dynamics of $^{48}\mathrm{Ca}$ levels and
  \ensuremath{\gamma}-ray decays from neutron inelastic scattering},\ }\href
  {https://doi.org/10.1103/PhysRevC.45.1628} {\bibfield  {journal} {\bibinfo
  {journal} {Phys. Rev. C}\ }\textbf {\bibinfo {volume} {45}},\ \bibinfo
  {pages} {1628} (\bibinfo {year} {1992})}\BibitemShut {NoStop}%
\bibitem [{\citenamefont {Speidel}\ \emph {et~al.}(2008)\citenamefont
  {Speidel}, \citenamefont {Schielke}, \citenamefont {Leske}, \citenamefont
  {Pietralla}, \citenamefont {Ahn}, \citenamefont {Costin}, \citenamefont
  {Zell}, \citenamefont {Gerber}, \citenamefont {Maier-Komor}, \citenamefont
  {Robinson}, \citenamefont {Escuderos}, \citenamefont {Sharon},\ and\
  \citenamefont {Zamick}}]{40Ar_Speidel_2008}%
  \BibitemOpen
  \bibfield  {author} {\bibinfo {author} {\bibfnamefont {K.-H.}\ \bibnamefont
  {Speidel}}, \bibinfo {author} {\bibfnamefont {S.}~\bibnamefont {Schielke}},
  \bibinfo {author} {\bibfnamefont {J.}~\bibnamefont {Leske}}, \bibinfo
  {author} {\bibfnamefont {N.}~\bibnamefont {Pietralla}}, \bibinfo {author}
  {\bibfnamefont {T.}~\bibnamefont {Ahn}}, \bibinfo {author} {\bibfnamefont
  {A.}~\bibnamefont {Costin}}, \bibinfo {author} {\bibfnamefont
  {O.}~\bibnamefont {Zell}}, \bibinfo {author} {\bibfnamefont {J.}~\bibnamefont
  {Gerber}}, \bibinfo {author} {\bibfnamefont {P.}~\bibnamefont {Maier-Komor}},
  \bibinfo {author} {\bibfnamefont {S.~J.~Q.}\ \bibnamefont {Robinson}},
  \bibinfo {author} {\bibfnamefont {A.}~\bibnamefont {Escuderos}}, \bibinfo
  {author} {\bibfnamefont {Y.~Y.}\ \bibnamefont {Sharon}},\ and\ \bibinfo
  {author} {\bibfnamefont {L.}~\bibnamefont {Zamick}},\ }\bibfield  {title}
  {\bibinfo {title} {New shell model calculations for $^{40}\mathrm{Ar}$ based
  on recent $g$-factor and lifetime measurements},\ }\href
  {https://doi.org/10.1103/PhysRevC.78.017304} {\bibfield  {journal} {\bibinfo
  {journal} {Phys. Rev. C}\ }\textbf {\bibinfo {volume} {78}},\ \bibinfo
  {pages} {017304} (\bibinfo {year} {2008})}\BibitemShut {NoStop}%
\bibitem [{\citenamefont {Bhattacharyya}\ \emph {et~al.}(2008)\citenamefont
  {Bhattacharyya}, \citenamefont {Rejmund}, \citenamefont {Navin},
  \citenamefont {Caurier}, \citenamefont {Nowacki}, \citenamefont {Poves},
  \citenamefont {Chapman}, \citenamefont {O'Donnell}, \citenamefont {Gelin},
  \citenamefont {Hodsdon}, \citenamefont {Liang}, \citenamefont {Mittig},
  \citenamefont {Mukherjee}, \citenamefont {Rejmund}, \citenamefont {Rousseau},
  \citenamefont {Roussel-Chomaz}, \citenamefont {Spohr},\ and\ \citenamefont
  {Theisen}}]{48Ar_Sarmi_PRL}%
  \BibitemOpen
  \bibfield  {author} {\bibinfo {author} {\bibfnamefont {S.}~\bibnamefont
  {Bhattacharyya}}, \bibinfo {author} {\bibfnamefont {M.}~\bibnamefont
  {Rejmund}}, \bibinfo {author} {\bibfnamefont {A.}~\bibnamefont {Navin}},
  \bibinfo {author} {\bibfnamefont {E.}~\bibnamefont {Caurier}}, \bibinfo
  {author} {\bibfnamefont {F.}~\bibnamefont {Nowacki}}, \bibinfo {author}
  {\bibfnamefont {A.}~\bibnamefont {Poves}}, \bibinfo {author} {\bibfnamefont
  {R.}~\bibnamefont {Chapman}}, \bibinfo {author} {\bibfnamefont
  {D.}~\bibnamefont {O'Donnell}}, \bibinfo {author} {\bibfnamefont
  {M.}~\bibnamefont {Gelin}}, \bibinfo {author} {\bibfnamefont
  {A.}~\bibnamefont {Hodsdon}}, \bibinfo {author} {\bibfnamefont
  {X.}~\bibnamefont {Liang}}, \bibinfo {author} {\bibfnamefont
  {W.}~\bibnamefont {Mittig}}, \bibinfo {author} {\bibfnamefont
  {G.}~\bibnamefont {Mukherjee}}, \bibinfo {author} {\bibfnamefont
  {F.}~\bibnamefont {Rejmund}}, \bibinfo {author} {\bibfnamefont
  {M.}~\bibnamefont {Rousseau}}, \bibinfo {author} {\bibfnamefont
  {P.}~\bibnamefont {Roussel-Chomaz}}, \bibinfo {author} {\bibfnamefont
  {K.-M.}\ \bibnamefont {Spohr}},\ and\ \bibinfo {author} {\bibfnamefont
  {C.}~\bibnamefont {Theisen}},\ }\bibfield  {title} {\bibinfo {title}
  {Structure of neutron-rich ar isotopes beyond $n=28$},\ }\href
  {https://doi.org/10.1103/PhysRevLett.101.032501} {\bibfield  {journal}
  {\bibinfo  {journal} {Phys. Rev. Lett.}\ }\textbf {\bibinfo {volume} {101}},\
  \bibinfo {pages} {032501} (\bibinfo {year} {2008})}\BibitemShut {NoStop}%
\bibitem [{\citenamefont {Peng}\ \emph {et~al.}(1979)\citenamefont {Peng},
  \citenamefont {Stein}, \citenamefont {Sunier}, \citenamefont {Drake},
  \citenamefont {Moses}, \citenamefont {Cizewski},\ and\ \citenamefont
  {Tesmer}}]{44-46Ca_Peng_PRL}%
  \BibitemOpen
  \bibfield  {author} {\bibinfo {author} {\bibfnamefont {J.~C.}\ \bibnamefont
  {Peng}}, \bibinfo {author} {\bibfnamefont {N.}~\bibnamefont {Stein}},
  \bibinfo {author} {\bibfnamefont {J.~W.}\ \bibnamefont {Sunier}}, \bibinfo
  {author} {\bibfnamefont {D.~M.}\ \bibnamefont {Drake}}, \bibinfo {author}
  {\bibfnamefont {J.~D.}\ \bibnamefont {Moses}}, \bibinfo {author}
  {\bibfnamefont {J.~A.}\ \bibnamefont {Cizewski}},\ and\ \bibinfo {author}
  {\bibfnamefont {J.~R.}\ \bibnamefont {Tesmer}},\ }\bibfield  {title}
  {\bibinfo {title} {Study of the reactions
  $^{46,48}\mathrm{Ti}$($^{14}\mathrm{C}$,
  $^{16}\mathrm{O}$)$^{44,46}\mathrm{Ca}$ and
  $^{50,52}\mathrm{Cr}$($^{14}\mathrm{C}$,
  $^{16}\mathrm{O}$)$^{48,50}\mathrm{Ti}$ at 51 mev},\ }\href
  {https://doi.org/10.1103/PhysRevLett.43.675} {\bibfield  {journal} {\bibinfo
  {journal} {Phys. Rev. Lett.}\ }\textbf {\bibinfo {volume} {43}},\ \bibinfo
  {pages} {675} (\bibinfo {year} {1979})}\BibitemShut {NoStop}%
\bibitem [{\citenamefont {Ideguchi}\ \emph {et~al.}(2010)\citenamefont
  {Ideguchi}, \citenamefont {Ota}, \citenamefont {Morikawa}, \citenamefont
  {Oshima}, \citenamefont {Koizumi}, \citenamefont {Toh}, \citenamefont
  {Kimura}, \citenamefont {Harada}, \citenamefont {Furutaka}, \citenamefont
  {Nakamura}, \citenamefont {Kitatani}, \citenamefont {Hatsukawa},
  \citenamefont {Shizuma}, \citenamefont {Sugawara}, \citenamefont {Miyatake},
  \citenamefont {Watanabe}, \citenamefont {Hirayama},\ and\ \citenamefont
  {Oi}}]{40Ar_Ideguchi_sd_band2010}%
  \BibitemOpen
  \bibfield  {author} {\bibinfo {author} {\bibfnamefont {E.}~\bibnamefont
  {Ideguchi}}, \bibinfo {author} {\bibfnamefont {S.}~\bibnamefont {Ota}},
  \bibinfo {author} {\bibfnamefont {T.}~\bibnamefont {Morikawa}}, \bibinfo
  {author} {\bibfnamefont {M.}~\bibnamefont {Oshima}}, \bibinfo {author}
  {\bibfnamefont {M.}~\bibnamefont {Koizumi}}, \bibinfo {author} {\bibfnamefont
  {Y.}~\bibnamefont {Toh}}, \bibinfo {author} {\bibfnamefont {A.}~\bibnamefont
  {Kimura}}, \bibinfo {author} {\bibfnamefont {H.}~\bibnamefont {Harada}},
  \bibinfo {author} {\bibfnamefont {K.}~\bibnamefont {Furutaka}}, \bibinfo
  {author} {\bibfnamefont {S.}~\bibnamefont {Nakamura}}, \bibinfo {author}
  {\bibfnamefont {F.}~\bibnamefont {Kitatani}}, \bibinfo {author}
  {\bibfnamefont {Y.}~\bibnamefont {Hatsukawa}}, \bibinfo {author}
  {\bibfnamefont {T.}~\bibnamefont {Shizuma}}, \bibinfo {author} {\bibfnamefont
  {M.}~\bibnamefont {Sugawara}}, \bibinfo {author} {\bibfnamefont
  {H.}~\bibnamefont {Miyatake}}, \bibinfo {author} {\bibfnamefont
  {Y.}~\bibnamefont {Watanabe}}, \bibinfo {author} {\bibfnamefont
  {Y.}~\bibnamefont {Hirayama}},\ and\ \bibinfo {author} {\bibfnamefont
  {M.}~\bibnamefont {Oi}},\ }\bibfield  {title} {\bibinfo {title}
  {Superdeformation in asymmetric n>z nucleus 40ar},\ }\href
  {https://doi.org/https://doi.org/10.1016/j.physletb.2010.02.031} {\bibfield
  {journal} {\bibinfo  {journal} {Physics Letters B}\ }\textbf {\bibinfo
  {volume} {686}},\ \bibinfo {pages} {18} (\bibinfo {year} {2010})}\BibitemShut
  {NoStop}%
\end{thebibliography}

\providecommand{\noopsort}[1]{}\providecommand{\singleletter}[1]{#1}%
%


\end{document}